\documentclass[pdflatex,sn_mathphys_num,iicol]{sn_jnl}

\usepackage[numbers,compress]{natbib}
\usepackage{graphicx}%
\usepackage{multirow}%
\usepackage{amsmath,amssymb,amsfonts}%
\usepackage{amsthm}%
\usepackage{mathrsfs}%
\usepackage[title]{appendix}%
\usepackage{xcolor}%
\usepackage{textcomp}%
\usepackage{manyfoot}%
\usepackage{booktabs}%
\usepackage{algorithm}%
\usepackage{algorithmicx}%
\usepackage{algpseudocode}%
\usepackage{listings}%
\usepackage{siunitx} 
\newcommand{\un}[1]{\, \unit{#1}} 
\newcommand{\hex}{_\mathrm{HEX}} 

\theoremstyle{thmstyleone}%
\theoremstyle{thmstyletwo}%

\theoremstyle{thmstylethree}%

\begin{document}

\title[Article Title]{QKD protected fiber-based infrastructure for time dissemination}


\author[1]{\fnm{Alice} \sur{Meda}}\email{a.meda@inrim.it}
\author*[1]{\fnm{Alberto} \sur{Mura}}\email{a.mura@inrim.it}
\author*[1]{\fnm{Salvatore} \sur{Virzì}}\email{s.virzi@inrim.it}
\author[1]{\fnm{Alessio} \sur{Avella}}\email{a.avella@inrim.it}
\author[1]{\fnm{Filippo} \sur{Levi}}\email{f.levi@inrim.it}
\author[1]{\fnm{Ivo} \sur{Degiovanni}}\email{i.degiovanni@inrim.it}
\author[2]{\fnm{Andrea} \sur{Geraldi}}\email{andrea.geraldi@thalesaleniaspace.com}
\author[2]{\fnm{Mauro} \sur{Valeri}}\email{mauro.valeri-somministrato@thalesaleniaspace.com}
\author[2]{\fnm{Silvia} \sur{Di Bartolo}}\email{silvia.dibartolo@thalesaleniaspace.com}
\author[2]{\fnm{Tommaso} \sur{Catuogno}}\email{tommaso.catuogno@thalesaleniaspace.com}
\author[2]{\fnm{Mattia} \sur{Verducci}}\email{mattia.verducci@thalesaleniaspace.com}
\author[1,3]{\fnm{Marco} \sur{Genovese}}\email{m.genovese@inrim.it}
\author[1]{\fnm{Davide} \sur{Calonico}}\email{d.calonico@inrim.it}


\affil[1]{
\orgname{Istituto Nazionale di Ricerca Metrologica (INRiM)}, \orgaddress{\street{Strada delle Cacce, 91}, \city{Torino}, \postcode{10135}, \country{Italy}}}

\affil[2]{ \orgname{Thales Alenia Space Italia}, \orgaddress{\street{Via Saccomuro, 24}, \city{Roma}, \postcode{00131}, \country{Italy}}}

\affil[3]{\orgdiv{INFN}, \orgname{Sez. di Torino
}, \orgaddress{\street{Via P. Giuria, 1}, \city{Torino}, \postcode{10125}, \country{Italy}}}


\abstract{
In this study, we demonstrate the possibility to protect, with Quantum Key Distribution (QKD), a critical infrastructure as the fiber-based one used for time and frequency (TF) dissemination service. The proposed technique allows to disseminate secure and precise TF signals between two fiber-optic-connected locations, on a critical infrastructure, using both QKD and White Rabbit technique.  This secure exchange allows the secret sharing of time information between two parties for the synchronization of distant clocks with the highest stability and traceable to the Italian time scale. 
When encrypted, time signals would reveal to a third party no useful information about the synchronization status, providing a time stability two orders of magnitude worsened. 
}




\maketitle
\section{Introduction}\label{sec3}
Precise clock synchronization signal distribution is of the utmost importance in several applications (financial transactions, classical communications, research activities, etc.) \cite{Lew99,And98} and the infrastructures for accurate time dissemination are considered as a part of the critical digital infrastructures that our society must protect  \cite{Enisa}.  
In the last years, the need to make critical digital infrastructures resilient to cyberattacks has been increased.  One of the challenges is to establish network security safe against the development of the quantum computer (QC) that is rapidly scaling up the number of qubits \cite{Lad10}. QC could potentially break current cryptographic systems, since their security is related only to
computational complexity \cite{Riv78,Ber17}. 

Quantum key distribution (QKD) is nowadays the most mature technology for sharing cryptographic secret keys with a level of security independent on computational power \cite{BB84, Sca09, Lo14, Rib24}. 
The consideration of QKD systems and networks from use cases has already started \cite{Lew22, Cic22,Wri21,Lop19, Pic23} and QKD metropolitan networks have been demonstrated all over the world \cite{Mir10, Pee09, Stu11}; in the UK, metropolitan quantum networks have been built by the Quantum Communications Hub in Cambridge and Bristol, connected by a long connection passing from London \cite{Xia23}. Quantum digital signatures were demonstrated in the NICT metro network in Tokyo \cite{Sas11}. In China, a $2000\un{Km}$ backbone connects Beijing and Shanghai, supplemented by ground-satellite QKD that exploits Micius satellite to extend QKD to global distances \cite{China, Wang14}.

The European Union (EU) is strongly pushing in this direction by signing with all the 27 member states a declaration to work together for the realization of the European Quantum Communication Infrastructure (EuroQCI) \cite{EQCI1}, that aims to realize quantum communication networks in all the EU countries. Within EuroQCI initiative,  the italian project QUID (Quantum Italy Deployment) \cite{QUID} proposes to start the deployment of the Italian part of the EU network by spreading QKD systems and networks in different cities, realizing several Quantum Metropolitan Area networks (QMANs). QMANs will be all connected by common fiber backbone, the Italian Quantum Backbone (IQB). This infrastructure was realized by the Italian Metrological Institute INRiM \cite{Clivati:20} and hosts the fiber-based dissemination service of an optical ultra-stable radiation referenced to the national primary frequency standard, which has an accuracy of $2 \cdot 10^{-16}$ \cite{Sta1}, and implements the Italian time transfer using the White Rabbit technique \cite{Lip11}, an improved version of the Precise Time Protocol (PTP) first defined by IEEE Standard 1588-2008 \cite{Sta1} (see Supplemental Material).
PTP technique can achieve accuracy of the order of few hundred nanoseconds. White Rabbit PTP (WR-PTP) was developed at CERN and incorporated in the PTP revision  (IEEE 1588 2019 \cite{Sta2}): it allows synchronizing distant clocks at nanosecond level or even better in properly designed and calibrated network architectures. 
 For critical applications, the protection of such infrastructures by sharing encrypted WR-PTP signals between two parties might result necessary to avoid attacks tailored on the clock synchronization or time information between parties.  
However, the WR-PTP is a standard protocol not designed for being easily encrypted. Furthermore, the common electronics devices for WR-PTP transmissions do not easily allow for internal modification of their input or output signals. 

Here we show an easy and reliable technique for encrypting WR-PTP transmissions by using QKD, realizing the QKD-protected time dissemination service exploiting the emergent Italian EuroQCI network. The sharing of the same infrastructure for precise time dissemination and QKD services open the way to  unique opportunity to simultaneously protect and distribute time information. 

\begin{figure}[t]
	\includegraphics[width=1\columnwidth]{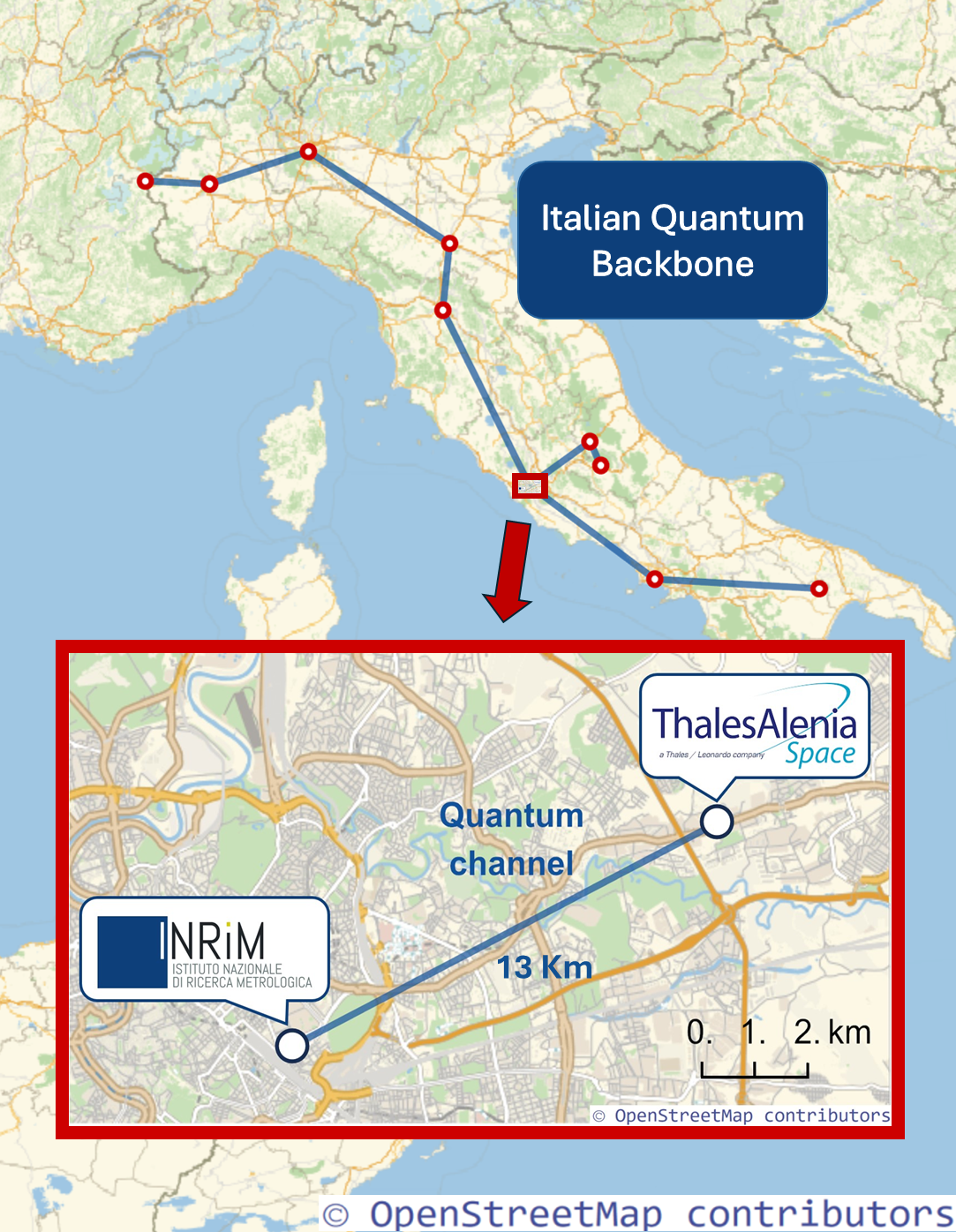}
	\caption{Schematic representation of the Italian Quantum Backbone showing the main links. The inset show the fiber link, inside the city of Rome, on which our technique was experimentally carried out.}
	\label{Map}
\end{figure}

To demonstrate the feasibility of our technique, we share encrypted WR-PTP in a real-world scenario between two nodes of the IQB, specifically in the Rome QMAN (see Fig. \ref{Map}) 

The work demonstrates the possibility to offer a quantum safe critical service  that can be easily implemented and extended to the entire IQB and to all the QMANs in Italy, providing a capillar dissemination of a QKD protected time service.  



\section{Results} 
\subsection{Encrypted WR-PTP}
We consider QKD and WR-PTP transmission in the end-to-end infrastructure that connects IQB nodes in INRiM and Thales Alenia Space Italy (TASI) premises, both in Rome. INRiM is the Alice (A) node, while Bob (B) node is in TASI. A and B host both QKD and WR-PTP equipment and signals are transmitted in the same optical fiber link.

\begin{figure}[htp!]
	\centering
	\includegraphics[width=1\columnwidth]{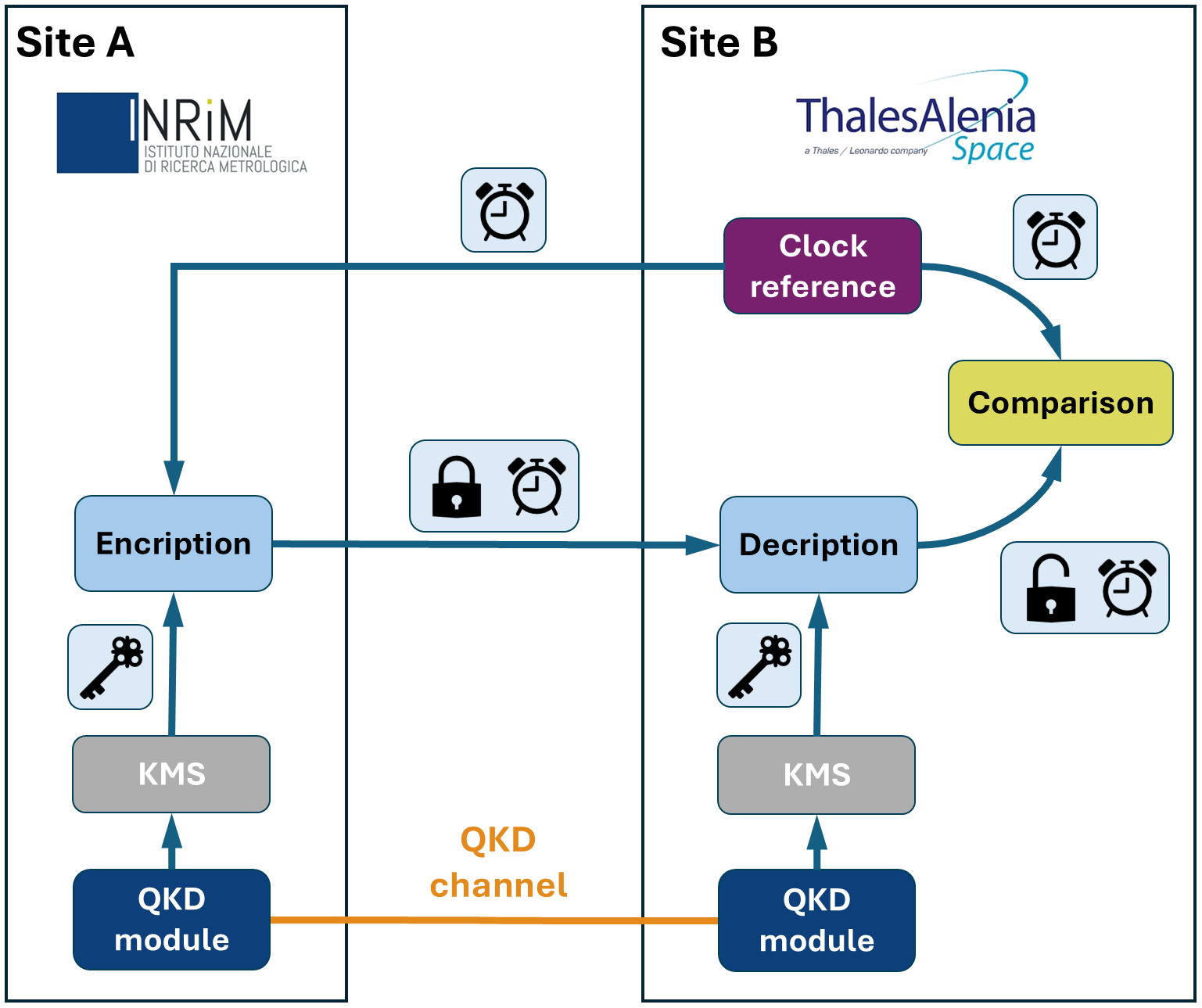}
	\caption{A schematic representation of the secure time dissemination experiment: a clock reference from Site B is sent to Site A via WR-PTP, encrypted with a QKD-generated key, and returned to Site B as encrypted WR-PTP. At Site B, the signal is decrypted and compared to the original clock reference for verification.}
	\label{schema}
\end{figure}

To demonstrate the feasibility of the secure time dissemination, we compare the signals from an high-performance external clock traceable to the Italian time scale UTC(IT), placed in B node, with itself after a round trip, where encryption in A and decryption in B with QKD occur (see Fig. \ref{schema}).

The clock reference
is an active hydrogen Maser clock (H-Maser) that provides a one pulse per second (1PPS) reference time signal and a frequency signal of $10\un{MHz}$.  
 During a standard WR-PTP transmission,  A and B share both the time and the frequency reference, coded by the WR-PTP protocol. 
 Sharing the same reference, it is possible to evaluate the real performances of the time encrypted dissemination.
 The A WR-PTP device, locked to an external frequency reference, converts phase variations of the frequency in temporal delays of the WR-PTP messages exchanged between A and B nodes. 
 If
 the frequency signal is shifted by a random phase, the time information signal in A and B will suffer of a random temporal delay $\Delta t$, making impossible for B to estimate the time information present in  A, and this is still true for an eventual eavesdropper.
We adopt an easy way to encrypt WR-PTP transmission, by using the quantum distributed keys between A and B to generate a secret truly random phase. A and B exchange keys and store them in a local Key Management System (KMS) in A and B. The user in A selects a key, generates the phase shift, applies it to the $10 \un{MHz}$ signal and sends the encrypted time signal to B; by using the same key, B is able to locally compensate the phase shift by inserting an opposite phase, obtaining the time signal, synchronized to the clock present in A.

\begin{figure*}[t!]
	\centering
	\includegraphics[width=2\columnwidth]{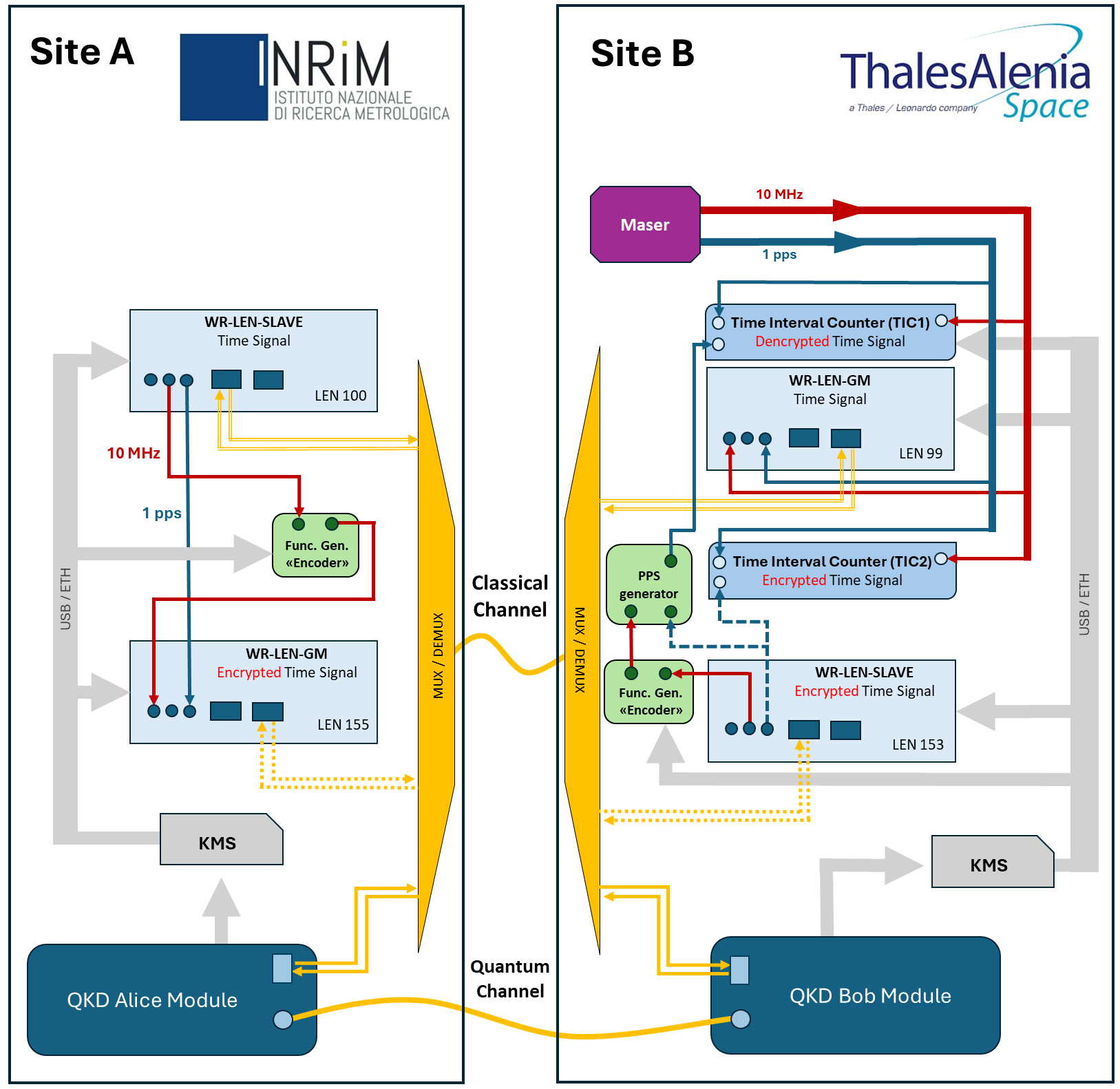}
	\caption{The experimental setup used for the encryption of the time signal with the QKD: at Site B, a maser generates a 10 MHz clock reference and a 1 PPS signal, which are transmitted to Site A via WR-PTP. At Site A, the WR-PTP signal is encrypted using a function generator with QKD-derived keys provided by the KMS. The encrypted WR-PTP signal is then sent back to Site B via the same classical channel. At Site B, the signal is decrypted using a corresponding function generator, and a PPS generator is exploited to realize 1PPs signals, measured with TIC1 and TIC2 for decrypted and encrypted WR-PTP signals, respectively. The WR-PTP signals exchange and the classical communication of the QKD devices are multiplexed over the same classical channel, while the QKD quantum channel is used to securely exchange encryption keys, which are managed by the KMS at both sites.}
	\label{setup}
\end{figure*}

We implement the phase shift starting from a white-noise model \cite{papoulis2002probability}. 

Since the shared key $K$ is exchanged as an array of random hexadecimal numbers, we organise $K$ in $N$ pairs:
\begin{equation}
	K = \{(k_1)\hex,\, \ldots ,\,(k_{2N})\hex\}
\end{equation}
 and for each pair we calculate the phase $\phi_i$:
\begin{equation}
	\phi_i=\frac{\left(k_{2i}k_{2i+1}\right)_{10}}{C} \,\,\,\,\,, \,\,\,\, i\in[0,N-1]
	\label{phi_i}
\end{equation}
where the notation $(x)_{10}$ indicates the hexadecimal number $x$ converted in decimal basis and $C$ is a constant integer number introduced because, without loosing of generality, it is experimentally convenient to limit the White phase noise added to a maximum level, below $2 \pi$.

\subsection{Experimental setup}

We implement our technique by using the experimental setup depicted in Fig. \ref{setup}, that shows the architecture for both the quantum key distribution and encrypted time exchange.


The INRiM station hosts the sender part of the QKD system, while the receiver is hosted in TASI location, connected through a fiber link for a total distance of $13\un{km}$. The link counts fibers where daily data traffic passes and two standard single-mode optical fibers dedicated to the experiment: one completely dedicated to the photons where the information on the key is encoded (Quantum channel) and one (Classical channel)  used to exchange WR-PTP signal, QKD synchronization and classical data for QKD post-processing, by means of multiplexer devices (MUX/DEMUX).
The Quantum channel needs a proper characterization to precisely estimate losses and level of background photons present in absence of QKD signal.
The QKD modules used to generate and share the keys are the IDQuantique Clavis$^3$ System, a discrete-variable QKD apparatus designed for research application that exploits two external single photon detectors to perform key exchange. In our experiment, we used two standard free-running InGaAs/InP Single-Photon Avalanche Diodes (SPADs), operating around $-50\un{^\circ C}$ with thermoelectric cooling, with  $20\un{\%}$ of detection efficiency and and dead time set to $25\un{\mu s}$.
The shared keys are stored and managed by means of local KMSs that can be access by users in A and B in a secure way.
 In order to test the technique, it is more convenient to use the H-maser present in B as the Master clock of the system. 
The Master clock of the experiment located in A is thus synchronized with a standard WR-PTP signal to the H-Maser.
By doing so it is easier to analyze the timing uncertainty of the encrypted (and decrypted) transmission from A to B. As a matter of fact, the location where we physically install the Master clock is an arbitrary choice for this experiment.


Our WR-PTP network is based on WR-LEN mini switches produced by Seven Solutions. These devices can be configured as Master or Slave \cite{Lip11} and allow to exchange WR-PTP traffic, as well as to set the 1PPS and $10\un{MHz}$ physical signals as input or output.




To enhance its stability and accuracy, the Master switch internal oscillator can be 
synchronized to the 1PPS and the $10\un{MHz}$ signal from an high-performance external clock.
When the Master is connected to another WR-LEN via the network interface, this Slave WR-LEN replicates output signals of 1PPS and $10\un{MHz}$ of the external high-performance clock connected to the Master. 

In our round trip, the raw 1PPS signal from the Master WR-PTP in B (WR-LEN-GM B), synchronized with the H-Maser, is sent to the WR Slave in A (WR-LEN-SLAVE A). 




A signal generator ("Encoder") varies the phase of the $10\un{MHz}$ signal of WR-LEN-SLAVE A with $\phi$, calculated from the key $K$ exchanged with QKD devices, according to Eq. \ref{phi_i}; to account for the limits of our instruments, we set $C = 4$, leading to a phase shift in the range  $\ang{0} \leq \phi_i \leq \ang{63.75}$, corresponding to a maximum temporal delay $\Delta t$ of about $17.71\un{ns}$.

Without loss of generality, for each interval $\Delta t$ we set the phase $-\phi_i$ in A node and $+\phi_i$ in B node.
The signal generator provides an input to the WR Master in A (WR-LEN-GM A). In this way, the WR-PTP signal that the Master in A exchanges with the Slave in B is phase shifted of $-\phi_i$; the result is that, after a round trip, the WR-LEN-SLAVE B clock (1PPS and $10\un{MHz}$ signals) is no more synchronized to UTC(IT), due to the phase shift periodically (every 5 seconds) and randomly changed.


Then, the $10\un{MHz}$ (encrypted) signal obtained from the WR-LEN-SLAVE B is sent to a second signal generator ("Decoder") that compensates the phase shift, allowing to recover the original time signal exploiting the PPS generator.
Both encrypted and decrypted signals are addressed to a Time Interval Counter (TIC): TIC1 and TIC2, respectively, for being compared to the original H-Maser time signal $t_0$.





\begin{figure}[htp!]
	\centering
	\includegraphics[width=1 \columnwidth]{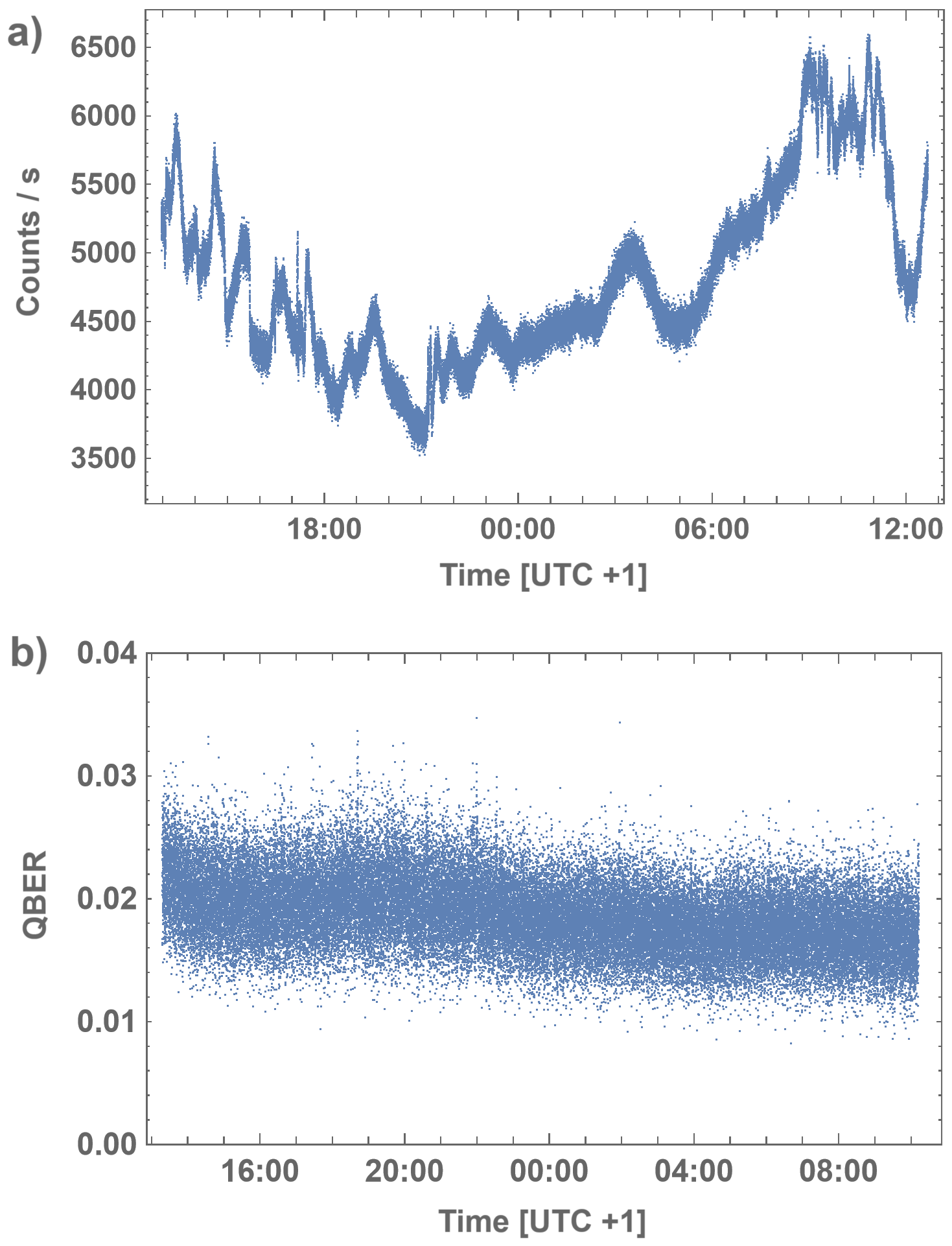}
	\caption{a) Background photons in a 24h acquisition frame. b) QBER as a function of time during a key exchange process}
	\label{QBER}
\end{figure}

 Before starting the QKD transmission,  we evaluated the presence of background photons in the Quantum fiber; as a matter of fact, 
Classical and Quantum fibers are part of a bundle of optical fibers, and the presence of unwanted photons due to evanescent coupling from the Classical channel and from data traffic from the other fibers must be evaluated. 
Thus, we measured for $24\un{h}$ the background photons present in the quantum fiber, monitoring any eventual behavior during all the day. We observed a spread distribution of about $(3500\,\div\,6500)\un{counts/s}$, probably due to the random data traffic. In our case, the intrinsic dark-count rate of the exploited SPAD is $353\un{counts/s}$, considerably less than the measured background photons.
However, such events are randomly distributed in time, whereas the protocol of our QKD devices works exploiting weak coherent pulses properly synchronized. Considering a transmission with repetition rate of $1\un{GHz}$, the background is always less than $6.5 \cdot 10^{-6}\un{counts/pulse}$ while the photons of the key are, considering the losses of our Quantum channel (about $10\un{dB}$), about $0.03\un{counts/pulse}$.  Moreover, since the sum of background and the QKD signal photons always remains below the saturation threshold due to the SPADs deadtime ($\sim 40\un{kcount/s}$), we can conclude that the channel is suitable for QKD transmission. We started the transmission, obtaining an exchange of keys with an average key rate of $1.5$ Kbit/s. The QKD devices were able to continuously exchange the keys with a low (about $2\un{\%}$) average Quantum Bit Error Rate (QBER). 
In Fig. \ref{QBER} we show both the background photons and the QBER behaviors for similar acquisition time.

\subsection{Measurement results}
Starting from the exchanged keys $K$, we compared both the encrypted and decrypted time signals to the original clock signal from the H-Maser with the two TICs, to directly observe the effect of the encryption (TIC2), and decryption (TIC1) with $K$.

\begin{figure}[htp!]
	\centering
	\includegraphics[width=1 \columnwidth]{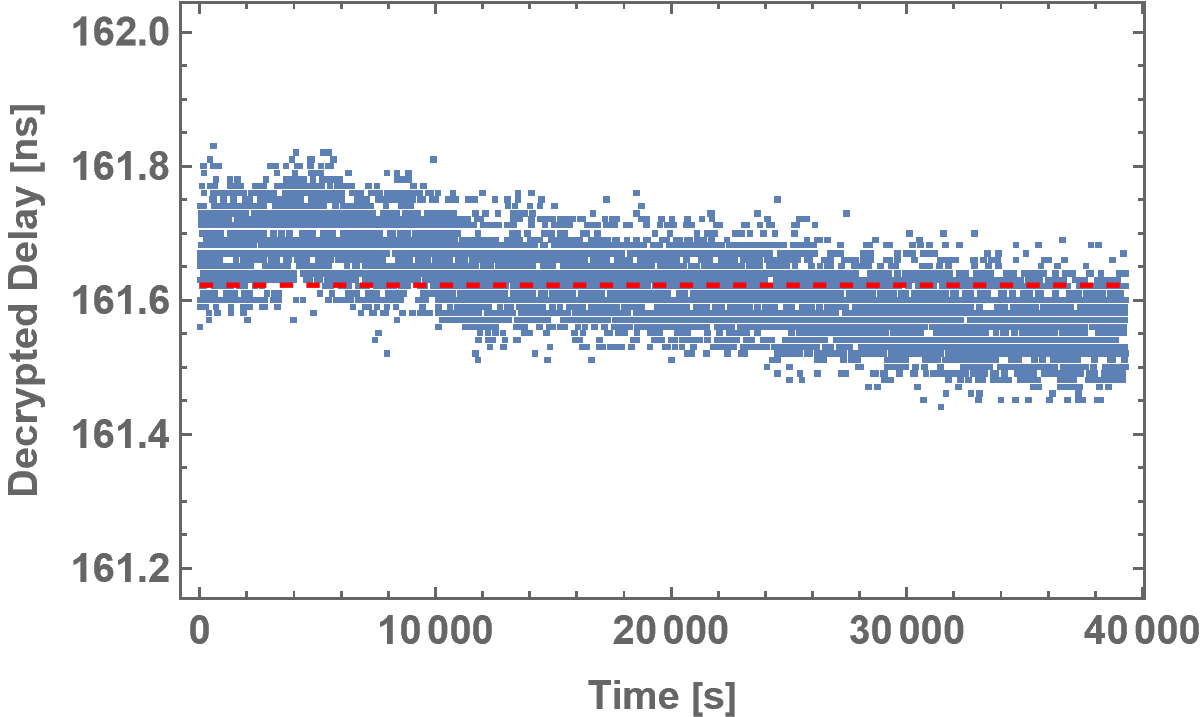}
	\caption{Decrypted TIC1 measured delay. The blue dots are the result of the measurement after the phase-decryption technique: the red dashed line represents the average value of such a delay. The acquisitions where repeated every $5\un{s}$, corresponding to the interval in which the phase remains constant. The system was monitored for more than $11\un{h}$.}
	\label{Phase_dec}
\end{figure}

\begin{figure}[t!]
	\includegraphics[width=1\columnwidth]{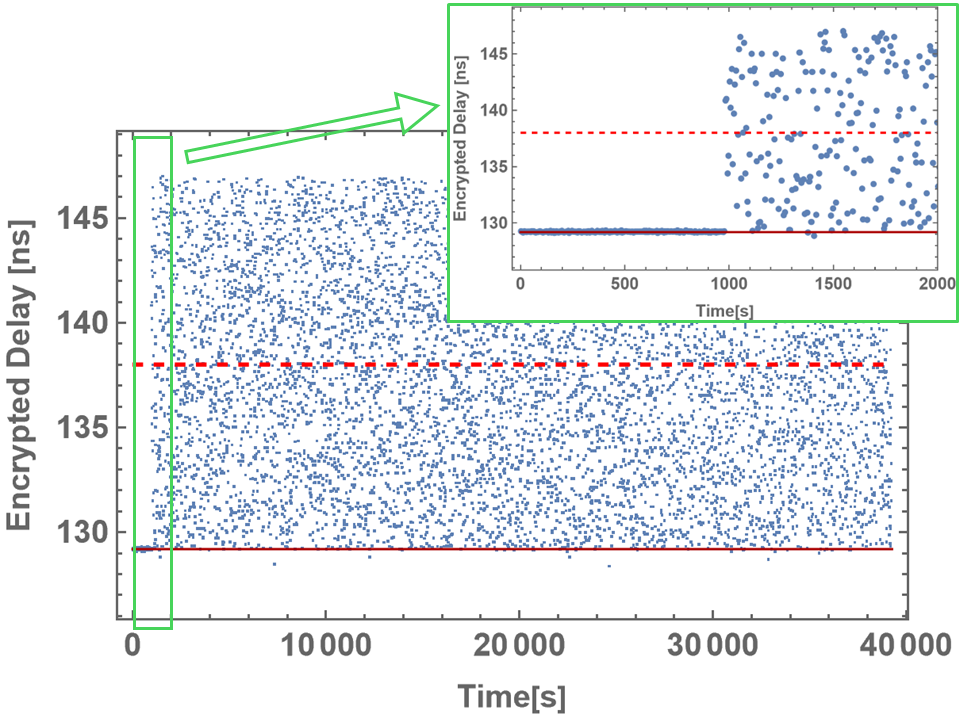}
	\caption{Encrypted TIC2 measured time information signal. The blue dots are the result of the measurement without decrypting the signal: the red dashed line represents the average value of such a delay. Initially, for about $15\un{min}$ Alice does not encrypted the WR signal, in order to estimate in the Bob side the real average delay, represented by the dark red line. The acquisitions where repeated every $5\un{s}$, corresponding to the interval in which the phase remains constant. The system was monitored for more than $11\un{h}$. The green inset highlights the behavior of the delay when Alice turns on the WR-signal encryption.}
	\label{Phase_enc}
\end{figure}

Fig.s \ref{Phase_dec},\ref{Phase_enc} show the results of the measured decrypted and encrypted delays, respectively. 
%

TIC1 measurements show the stability of the delay induced by the phase shift (blue dots) with respect to an average time delay (red dashed line) of $161.6224(7)\un{ns}$ during an acquisition time longer than $11\un{h}$.
On the contrary, TIC2 shows two different behaviors, highlighted into the green inset: a stable and a noisy one. This is due to the fact that we initially did not encrypted the transmission for about $15\un{min}$, in order to calibrate the real delay bias (dark red line) of $129.188(3)\un{ns}$. Subsequently, we turned on our encoding system, obtaining a spread distribution around a mean (red dashed line) of $138.00(6)\un{ns}$.
Both the phase shifting of the encrypted signal and the increasing of the dispersion of the time delay are evident, making the transmission not useful for a precise time synchronization. 

\begin{figure}[t!]
	\centering
	\includegraphics[width=1\columnwidth]{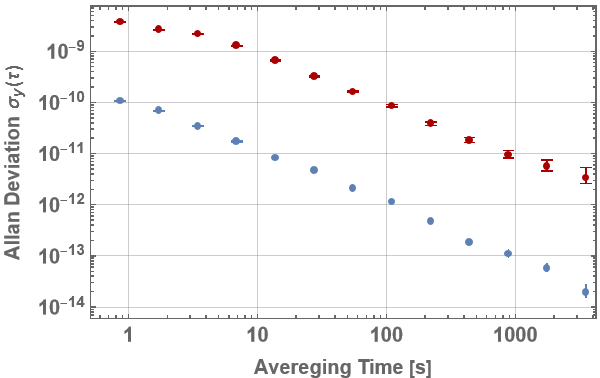}
	\caption{Frequency stability comparison between TIC1 and TIC2 measurements. They are plotted the Allan deviations measured in TIC1 (blue dots) and in TIC2 (red dots) depending on the averaging time with their corresponding uncertainties.}
	\label{allan_fig}
\end{figure}

Furthermore, we characterize the time delay stability of TIC1 and TIC2 measurements. Fig. \ref{allan_fig} shows the comparison between the Allan deviation of data from TIC1 (red) and TIC2 (blue) as a function of averaging time. In both cases one could recognize the linear behavior (in logarithmic scale) of the Allan deviation typical of white phase noise. However, our technique allowed for a degradation in the encrypted signal of more than two order of magnitude with respect to the decrypted one.

%

%
%

\section{Discussion}	
The results shown in Fig. \ref{Phase_enc} clearly demonstrate the robustness of our encoding technique, even for long time. Without a deterministic decryption, any measurement does not allow achieving the temporal precision usually granted exploiting the WR-PTP protocol. 
In addition, in Fig. \ref{allan_fig} it is possible to appreciate that our approach allows for gaining almost two order of magnitude in terms of Allan deviation with respect to an eventual eavesdropping. This means that, if a third-part is interested in the estimation of the Alice's clock frequency stability, our encoding technique would ensure that the information needs more time to be revealed, i.e. diminishing the eavesdropper possibility for extracting information.
 
For our proof-of-principle experiment, we used the most simple  
 noise model, i.e. white noise. However, although a more elaborated noise model could in principle grant for better results, specially in terms of Allan deviation, there is a physical limit on the allowed range for the introduced phase. Thus, taking into account a reasonable limit for the phase, e.g. $\phi_i \in (\ang{-360}\,,\,\ang{360})$, the behavior of a more complex noise will converge to a white noise in few steps. Of course, this is mediated by a trade-off between the average step size and the maximum allowed value of the phase. Simulating different noise models (even non-Markovian\cite{papoulis2002probability}) limited in the range $\phi_i \in (\ang{-360}\,,\,\ang{360})$ and maintaining the same average phase step implemented in our proof of principle, we obtained the convergence to a white-noise behavior in less than 10 steps (see the Supplemental Material).

In conclusion, our work demonstrates a robust technique to protect time signal with truly random noise, easily implementable in the quantum communication infrastructures that will host both time dissemination and QKD protocols, even with common electronics usually available in any laboratories. The synchronization information exhibits a stability at 1 second worsened by two orders of magnitude. This proof-of-principle experiment paves the way to a novel technique for quantum protection of a synchronization service, making timing signals not available to unauthorized users.

\section*{Data Availability}
All data needed to evaluate the conclusions in the paper are available from the corresponding author upon reasonable request.

\section*{Acknowledgments}
The results presented in this paper have been achieved in the context of the following projects: E2E Quantum Communication \& Synchronization TestBed
(QCS-TB), financed by Cyber 4.0 Bando 1-2021,  QUID (QUantum Italy
Deployment) and EQUO (European QUantum ecOsystems)
which are funded by the European Commission in the Digital Europe Programme under the grant agreements No 101091408
and 101091561; Qu-Test, which has received funding from
the European Union’s Horizon Europe, The EU Research
and Innovation Programme under the Grant Agreement number
101113983; ARS01 00734-QUANCOM (European structural
and investment funds MUR-PON Ricerca \& Innovazione
2014-2020); 22-EU-DIG-5G FREJUS, financed by CEF-DIG-2022-5GCORRIDORS-STUDIES under the Grant Agreement number 101133818

\section*{Author Contributions}
The preparation of the infrastructure, the experiment and the data analysis were run by A. Me., A. Mu., S. V., A. A. (principal investigators), A. G., M. V. and T. C., under the supervision of D. C., I.P.D., F. L., M. G. and S. D..

The manuscript was written with inputs from all the authors.

\section*{Competing Interests}
The authors declare no competing interests.

\bibliographystyle{unsrtnat}
\bibliography{Main_article}

\end{document}


\title[Article Title]{Supplemental Material}


\author[1]{\fnm{Alice} \sur{Meda}}\email{a.meda@inrim.it}
\author*[1]{\fnm{Alberto} \sur{Mura}}\email{a.mura@inrim.it}
\author*[1]{\fnm{Salvatore} \sur{Virzì}}\email{s.virzi@inrim.it}
\author[1]{\fnm{Alessio} \sur{Avella}}\email{a.avella@inrim.it}
\author[1]{\fnm{Filippo} \sur{Levi}}\email{f.levi@inrim.it}
\author[1]{\fnm{Ivo} \sur{Degiovanni}}\email{i.degiovanni@inrim.it}
\author[2]{\fnm{Andrea} \sur{Geraldi}}\email{andrea.geraldi@thalesaleniaspace.com}
\author[2]{\fnm{Mauro} \sur{Valeri}}\email{mauro.valeri-somministrato@thalesaleniaspace.com}
\author[2]{\fnm{Silvia} \sur{Di Bartolo}}\email{silvia.dibartolo@thalesaleniaspace.com}
\author[2]{\fnm{Tommaso} \sur{Catuogno}}\email{tommaso.catuogno@thalesaleniaspace.com}
\author[2]{\fnm{Mattia} \sur{Verducci}}\email{mattia.verducci@thalesaleniaspace.com}
\author[1,3]{\fnm{Marco} \sur{Genovese}}\email{m.genovese@inrim.it}
\author[1]{\fnm{Davide} \sur{Calonico}}\email{d.calonico@inrim.it}


\affil[1]{
\orgname{Istituto Nazionale di Ricerca Metrologica (INRIM)}, \orgaddress{\street{Strada delle Cacce, 91}, \city{Torino}, \postcode{10135}, \country{Italy}}}

\affil[2]{ \orgname{Thales Alenia Space Italia}, \orgaddress{\street{Via Saccomuro, 24}, \city{Roma}, \postcode{00131}, \country{Italy}}}

\affil[3]{\orgdiv{INFN}, \orgname{Sez. di Torino
}, \orgaddress{\street{Via P. Giuria, 1}, \city{Torino}, \postcode{10125}, \country{Italy}}}

 \keywords{QKD, White Rabbit, Time Dissemination}



    \maketitle

    \section*{A) Noise models}
    In this section, we present numerical simulations of various phase noise models, extending beyond the white noise\cite{papoulis2002probability} model described in the main text to progressively incorporate more complex correlations. 
    For the reader's convenience, the results of the white noise model used in the main experiment, shown in Fig.s 6, 7 of the main article, are reproduced in Fig. \ref{res:WN}. 
    \renewcommand{\thefigure}{SM1}
    \begin{figure}[h!]
    \centering
        \begin{subfigure}{0.49\textwidth}
        \includegraphics[width=\textwidth]{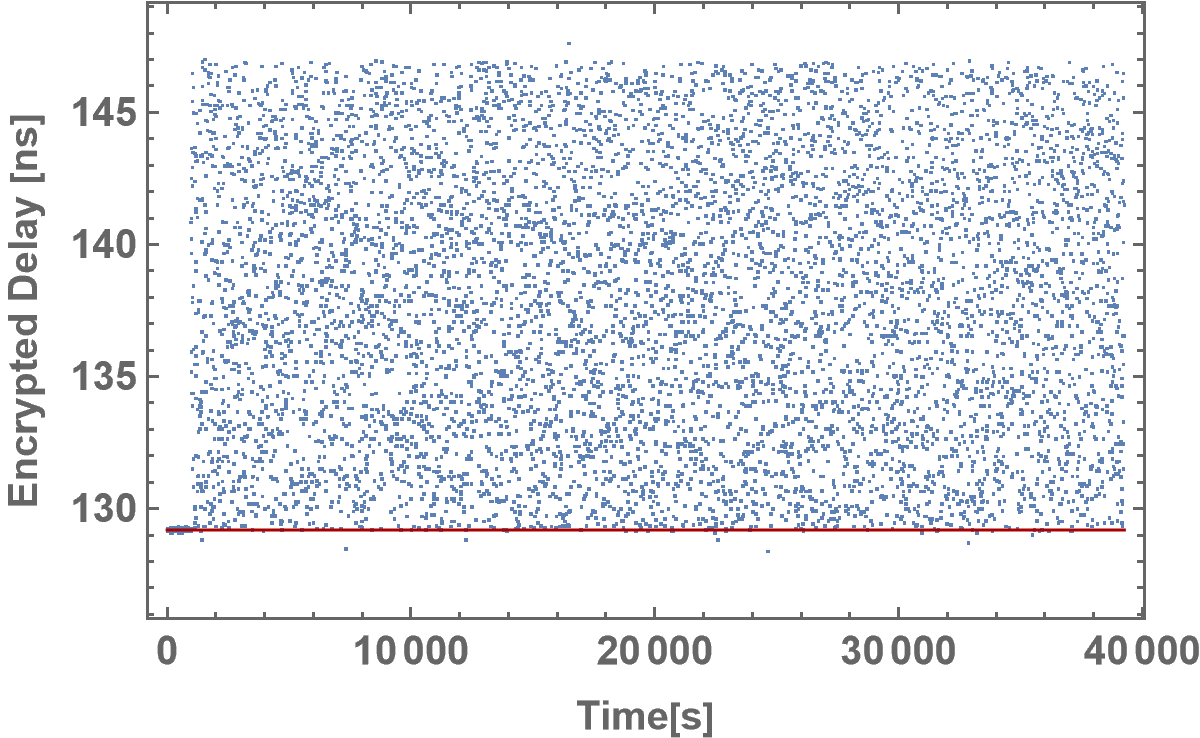}
        \caption{Time information signal}
        \label{A_res:WN}
    \end{subfigure}
        \begin{subfigure}{0.49\textwidth}
        \includegraphics[width=\textwidth]{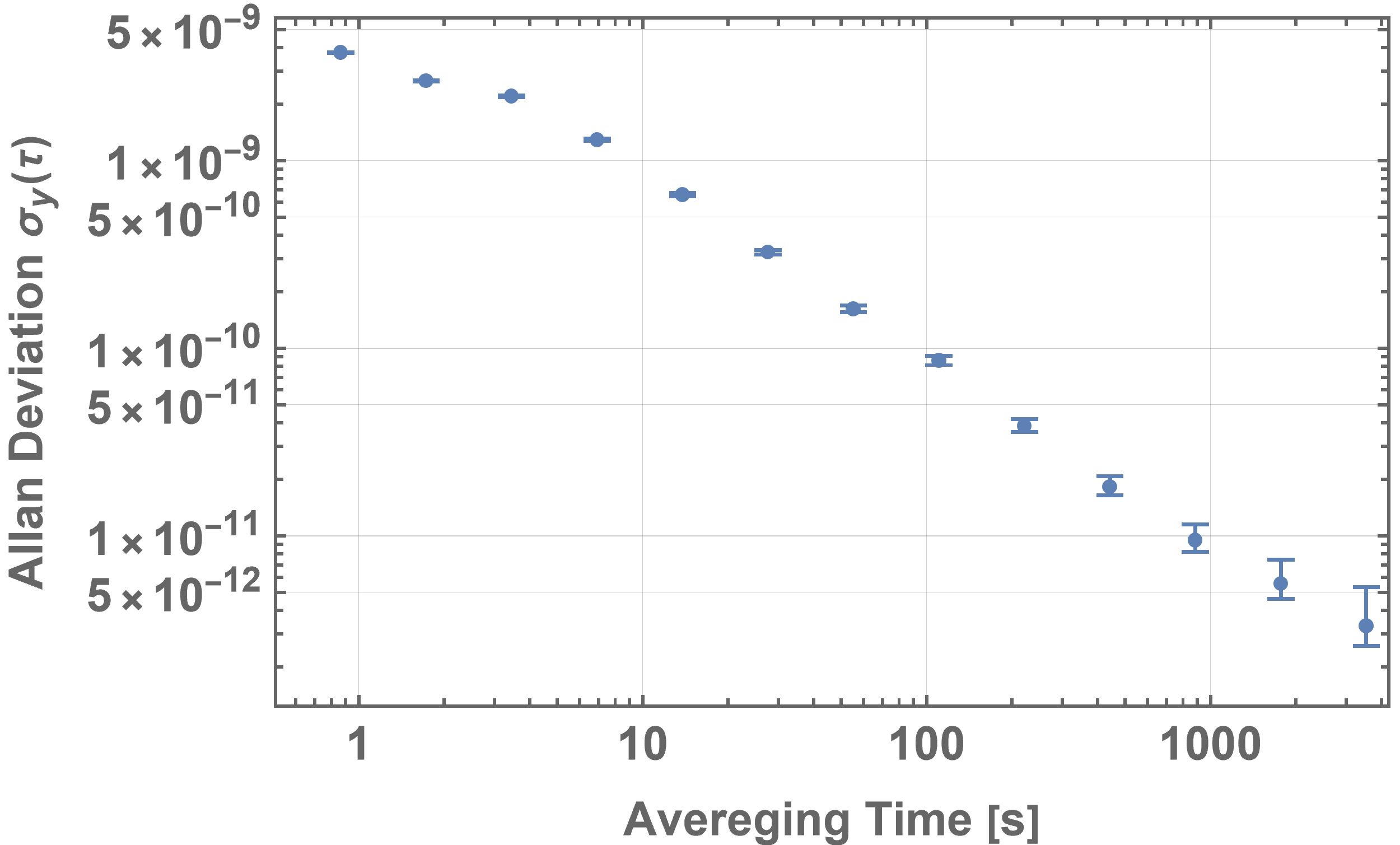}
        \caption{Frequency stability}
        \label{B_res:WN}
        \end{subfigure}        
    \caption{Results of the measurement performed during the experiment exploiting a white noise model. a) The blue dots are the result of the measurement without decryption, and the red line represents the average offset delay induced by electronics. b) Allan deviation of the measured time information signal depending on the averaging time with corresponding uncertainty. }
    \label{res:WN}
    \end{figure}
    The first model we discuss is the phase random walk (RW)\cite{papoulis2002probability}, which represents a simple yet fundamental step beyond white noise, characterized by an unbounded accumulation of phase deviations over time, resulting in a RW process.

    The second model introduces a non-Markovian\cite{papoulis2002probability} variant of the phase RW, where correlations are induced after a fixed number of $M$ steps. 
    This model introduces weak memory effects, as the phase deviations are no longer entirely independent but exhibit a degree of correlation after every $M$ steps. 

    Building upon the non-Markovian RW, the third model further extends the correlation structure. 
    In this model, the phase correlation at step $n+1$ depends on the preceding $S$ steps, offering a more intricate phase evolution influenced by multiple past states. 
    This model encapsulates a higher degree of memory effect, where the phase state is a cumulative result of a series of previous steps rather than being reset periodically. 

    Each of these models is simulated with and without imposing a phase limit, allowing us to observe the effects of a realistic range on phase behavior. 
    For each case, we present phase plots and Allan deviation graphs, providing a comprehensive analysis of the phase noise characteristics. 
    As mentioned in the main text, any model converges to a white-noise behavior within a few steps when a realistic phase range (in our case, $\ang{-360}$ to $\ang{360}$) is introduced. 
    This convergence highlights the impact of bounded phase limits on the overall noise characteristics, emphasizing the importance of considering such constraints in practical applications.

    \subsection*{Random Walk}
    To implement a RW behavior for the encrypting phase, we consider a portion of the exchanged secret key $K_3$ composed by $N$ hexadecimal triplets, each one giving the $i$-th step of the RW:
    \begin{equation}
        K_3 = \{ (k_1)\hex,\,\ldots ,\, (k_{3N})\hex \}
        \label{key}
        \tag{SM1}
    \end{equation}

    Starting from $K_3$, we would like to design the RW behavior as general as possible. 
    For each triplet, the first hexadecimal determines the sign of the step, while the remaining pair its magnitude.
    Therefore, the $i$-th phase step $\phi_i^{\mathrm{RW}}$ can be calculated as follows:
    \begin{equation}
        \phi_i^{\mathrm{RW}} = \left\{ 
        \begin{split}
            & \phi_{i-1}^{\mathrm{RW}} + \frac{\left(k_{3i+1}k_{3i+2}\right)_{10}}{C}\,\,\,\,\, \mathrm{if} \,\,\,\,\,(k_{3i})\hex \geq (T)\hex \\
            & \phi_{i-1}^{\mathrm{RW}} - \frac{\left(k_{3i+1}k_{3i+2}\right)_{10}}{C}\,\,\,\,\, \mathrm{if} \,\,\,\,\,(k_{3i})\hex < (T)\hex
        \end{split}
        \right. \,\,\,\,\,\,\,\,\, i \in [0,\,N-1]\,\,\,\,\,\quad\quad\quad
        \tag{SM2}
        \label{RW}
    \end{equation}
    where $\phi_{i-1}^{\mathrm{RW}}\arrowvert_{i=0} \equiv B$ is a fixed initial bias, $(T)\hex$ is a hexadecimal threshold and $C$ is a constant (see Eq. (2) in the main text). 
    In our case, we fixed $C = 4$ as in the proof of principle, and $(T)\hex = 8\hex$ (balanced RW).

    Without any constraint on the phase values, the simulation results are shown in Fig. \ref{res:RW}.
    \renewcommand{\thefigure}{SM2}
    \begin{figure}[h!]
    \centering
        \begin{subfigure}{0.49\textwidth}
        \includegraphics[width=\textwidth]{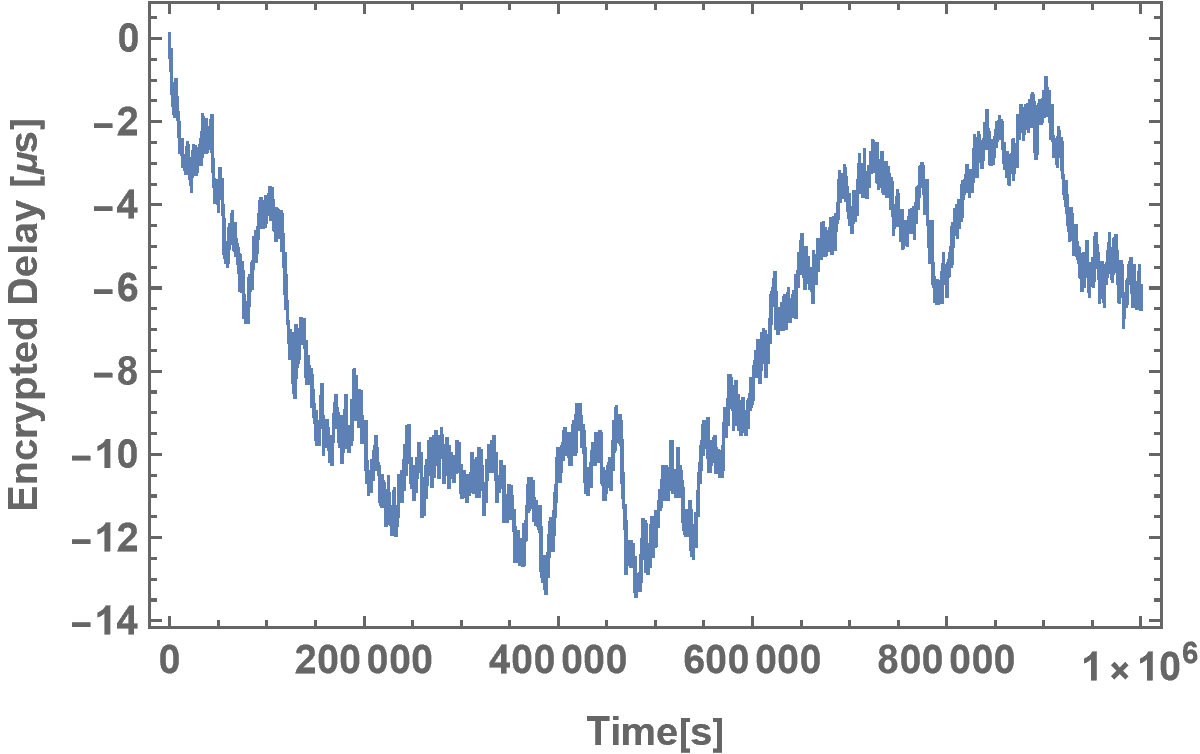}
        \caption{Time information signal}
        \label{A_res:RW}
    \end{subfigure}
        \begin{subfigure}{0.49\textwidth}
        \includegraphics[width=\textwidth]{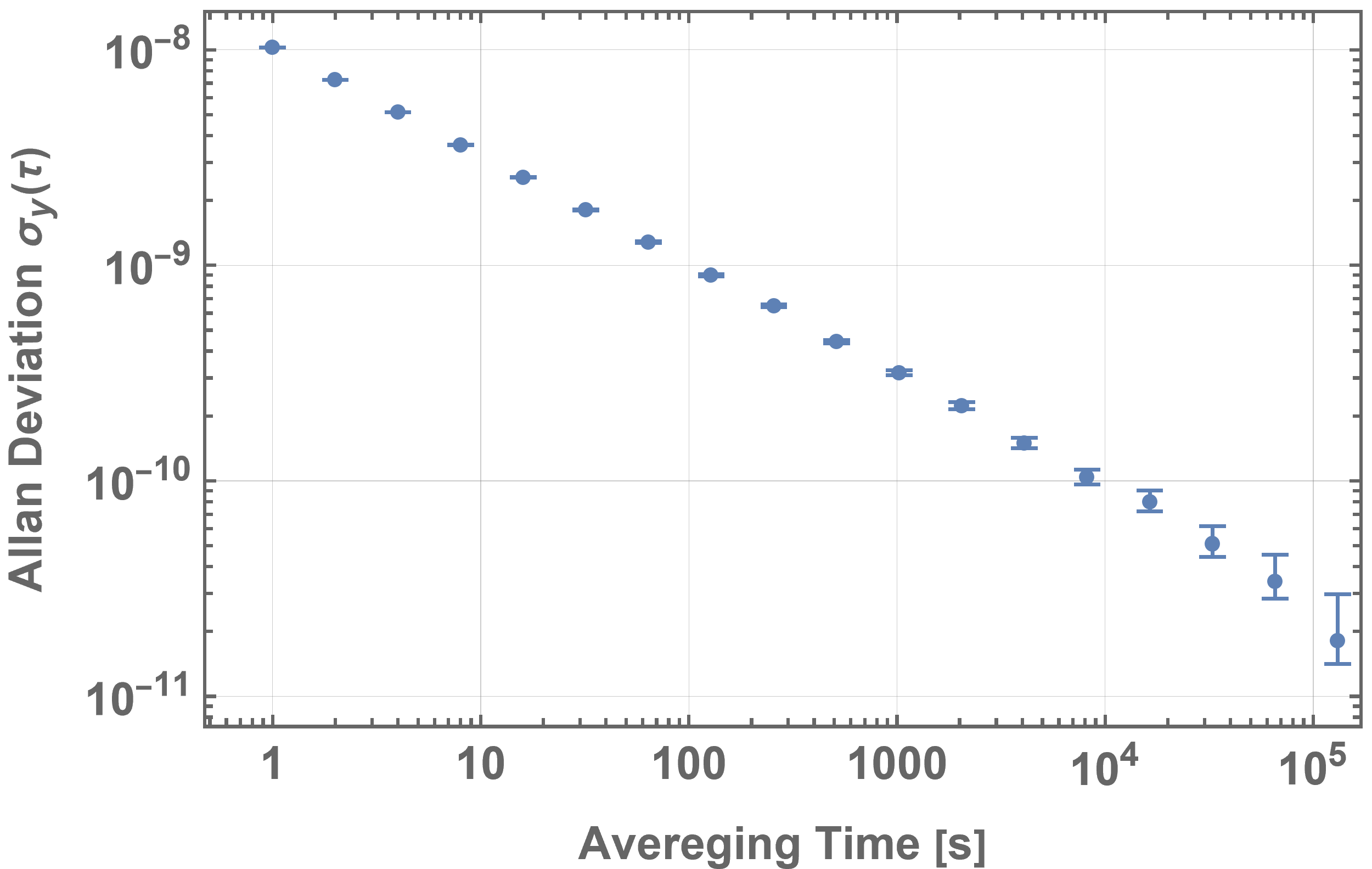}
        \caption{Frequency stability}
        \label{B_res:RW}
        \end{subfigure}        
    \caption{Simulation of the time information signal encrypted with a phase RW model. a) The blue dots are the result of the simulation without constraints for the phase range. b) Allan deviation of the simulated time information signal depending on the averaging time with corresponding uncertainty.}
    \label{res:RW}
    \end{figure}

    In Fig. \ref{A_res:RW}, we present the simulated encrypted time information signal, where the mean delay introduced by the electronics is assumed to be zero. 
    This idealized scenario allows us to isolate the effects of the phase noise models on the signal without the added complexity of electronic delays. 
    In Fig. \ref{B_res:RW}, we show the corresponding frequency stability, expressed through the Allan deviation, calculated from the data presented in Fig. \ref{A_res:RW}. 
    The frequency stability in Fig. \ref{B_res:RW} exhibits a characteristic $(1/\sqrt{\tau})$ behavior, which is notably advantageous compared to the $(1/\tau)$ trend observed in Fig. 7 of the main text, where the measurements were performed with white noise.

    The RW behavior depicted in Fig. \ref{B_res:RW} is particularly beneficial in the context of signal security. 
    This trend makes it more challenging for a potential eavesdropper to accurately estimate the stability of the clock used in the White Rabbit (WR) signal distribution. 
    The slower degradation of stability with increasing integration time ($\tau$) implies that the clock's stability is less predictable, adding an additional layer of security to the encrypted time distribution.

    However, when comparing these results to the acquisitions shown in Fig. 6 of the main text, the phase RW illustrated in Fig. \ref{A_res:RW} leads to a significantly larger delay in the encrypted time signal. 
    This delay, while substantial, corresponds to multiple cycles of the $10\un{MHz}$ signal delay, effectively rendering it impractical to maintain a WR transmission encoded with this method. 
    The extensive phase deviations make it impossible to reliably decode the transmitted signal without introducing errors, as the encrypted time information becomes highly ambiguous.

    To address this issue, we repeated the simulation by imposing a limit on the possible phase values according to:
     \begin{equation}
     \label{constraint}
        \overline{\phi_i^{\mathrm{RW}}} = R \sin\left( \phi_i^{\mathrm{RW}}\right)
        \tag{SM3}
    \end{equation}
    where $\phi_i^{\mathrm{RW}}$ is determined following Eq. \ref{RW}, and $R = \ang{360}$ does not allow for more than one period of induced delay.
    
    This constraint is necessary to ensure that the encrypted time signal remains within a manageable range, making it feasible to encode and decode the WR transmission effectively while maintaining the enhanced security provided by the phase noise model. 
    The results of this simulation are shown in Fig. \ref{res:RWlim}.
    \renewcommand{\thefigure}{SM3}
    \begin{figure}[H]
    \centering
        \begin{subfigure}{0.49\textwidth}
        \includegraphics[width=\textwidth]{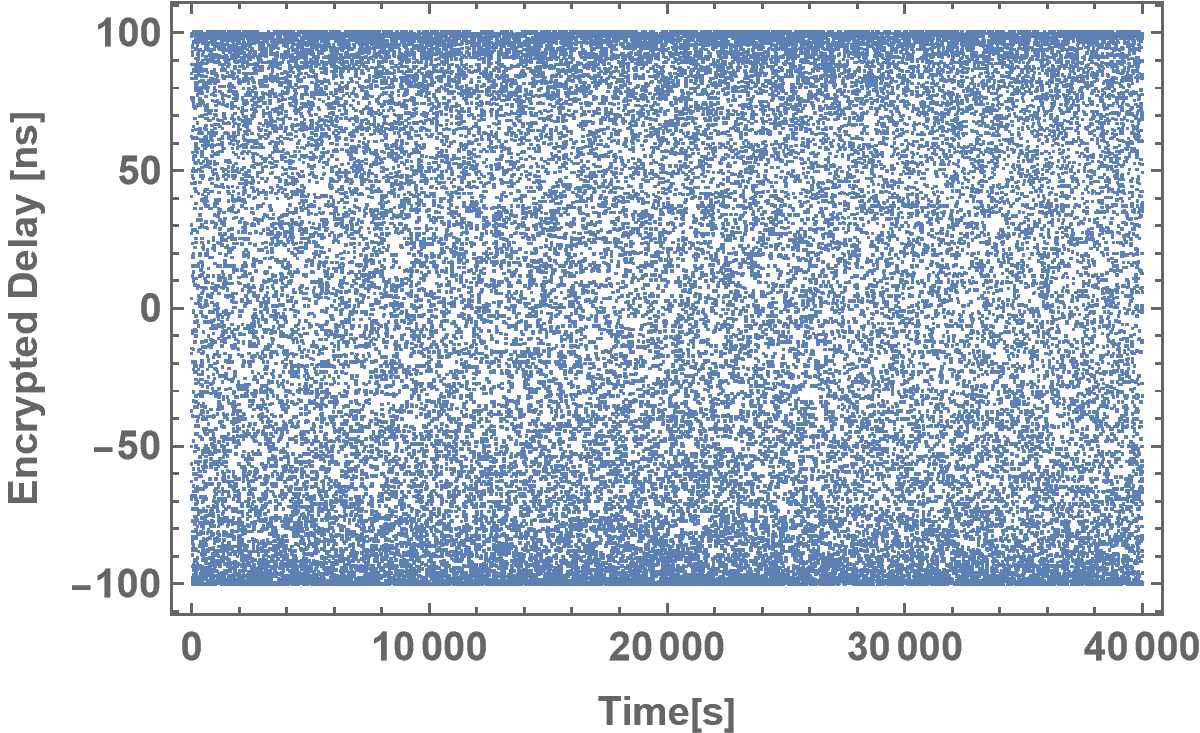}
        \caption{Time information signal}
        \label{A_res:RWlim}
    \end{subfigure}
        \begin{subfigure}{0.49\textwidth}
        \includegraphics[width=\textwidth]{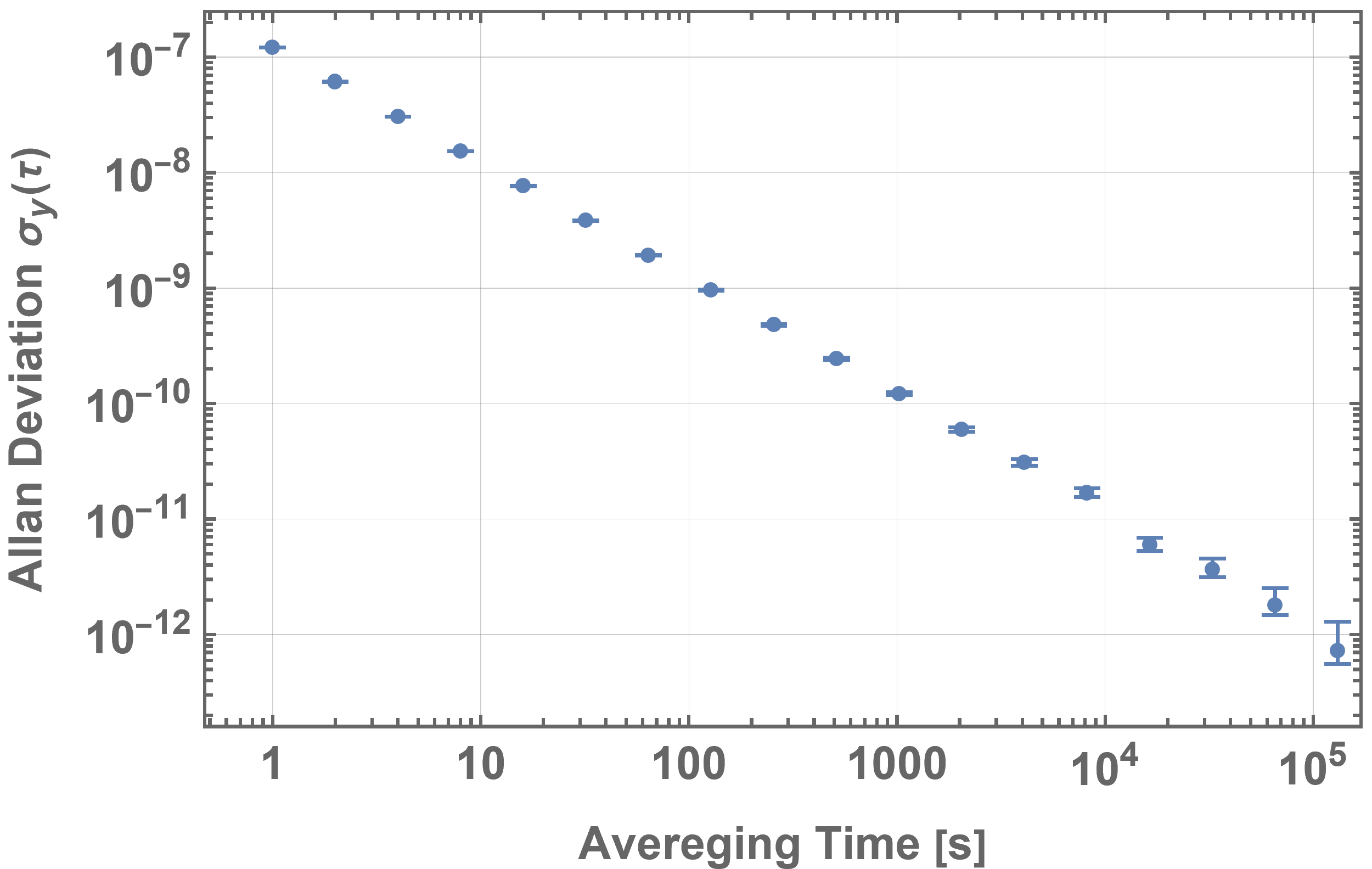}
        \caption{Frequency stability}
        \label{B_res:RWlim}
        \end{subfigure}        
    \caption{Results of the RW simulation with constraint on phase values according to Eq. \ref{constraint}. a) The blue dots represent the encrypted time information signal. b) Allan deviation of the simulated time information signal depending on the averaging time with corresponding uncertainty.}
    \label{res:RWlim}
    \end{figure}

    By introducing a necessary constraint for a real-world transmission, it becomes evident that Fig.s \ref{A_res:RWlim}, \ref{B_res:RWlim} closely resemble \ref{A_res:WN}, \ref{B_res:WN}, respectively.
    The differences between these figures are solely due to the range of possible phase values introduced. 
    In the experiment, this corresponds to a delay between $0\un{ns}$ and approximately $18\un{ns}$, whereas in the simulation, the delay range is as wide as $200\un{ns}$.
    
    Specifically, within a few steps, the phase values reach the imposed range bound, causing the system's evolution to mimic a white-noise behavior, thereby nullifying any advantages initially provided by the RW model. 
    One could consider increasing the value of the constant $C$ in Eq. \ref{RW} to decrease the step size of the phase RW, thereby delaying the phase from reaching its bound. 
    However, since the constraint on the range cannot realistically be expanded, this would require significantly reducing the step size, making the difference in terms of delay induced in the WR signal encoding negligible.

    \subsection*{Random Walk with correlation after $M$ steps}
    
    We now aim to introduce correlations into the RW model presented in Eq. \ref{RW}, thereby transforming it into a non-Markovian model. 
    In particular, starting from the key $K_3$, we introduced correlations in the model by setting at the $i$-th step a certain probability of repeating the sign of the step $i-M$, where $1<M<N-1$. Such a probability is given by the comparison of the first hexadecimal of the $i$-th key triplet (see Eq. \ref{key}) with respect to a threshold parameter $(T_M)\hex$.
    Then, the phase evolution model can be described as follows:
    \begin{equation}
    \phi_i^{\mathrm{RW}_M} = \left\{ 
        \begin{split}
            & \phi_{i}^{\mathrm{RW}}\quad\quad,\quad\quad i < M\\
            & \phi_{i-1}^{\mathrm{RW}_M} + \mathrm{sign}\left( \Delta\phi_{i-M}^{\mathrm{RW}_M} \right)\frac{\left(k_{3i+1}k_{3i+2}\right)_{10}}{C}\,\,\,\,\, \mathrm{if} \,\,\,\,\,(k_{3i})\hex \geq (T_M)\hex \,\wedge\, i>M\\
            & \phi_{i-1}^{\mathrm{RW}_M} - \mathrm{sign}\left( \Delta\phi_{i-M}^{\mathrm{RW}_M} \right)\frac{\left(k_{3i+1}k_{3i+2}\right)_{10}}{C}\,\,\,\,\, \mathrm{if} \,\,\,\,\,(k_{3i})\hex < (T_M)\hex  \,\wedge\, i>M
        \end{split}
    \right.\quad\quad\quad\,\,\,
    \tag{SM4}
    \label{RW_M}
    \end{equation}
    where $i \in \left[0,\, N-1\right]$, $\phi_{i}^{\mathrm{RW}}$ is defined in Eq. \ref{RW}, and $\Delta\phi_{i-M}^{\mathrm{RW}_M} \equiv \phi_{i-M}^{\mathrm{RW}_M} - \phi_{i-M-1}^{\mathrm{RW}_M}$. 
    In Fig. \ref{res:RW_M} we show the simulated results obtained setting $M=100$ in Eq. \ref{RW_M}, without putting any constraint to the resulting phase dynamics.
    \renewcommand{\thefigure}{SM4}
    \begin{figure}[H]
    \centering
        \begin{subfigure}{0.49\textwidth}
        \includegraphics[width=\textwidth]{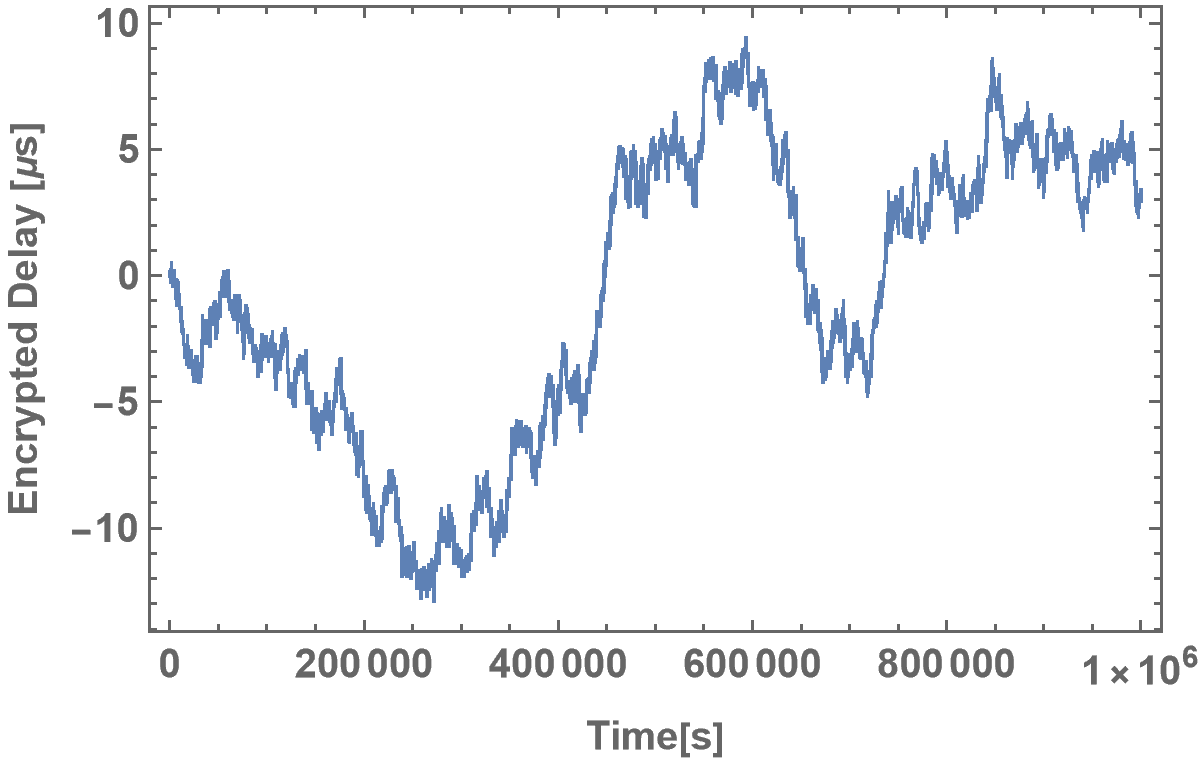}
        \caption{Time information signal}
        \label{A_res:RW_M}
    \end{subfigure}
        \begin{subfigure}{0.49\textwidth}
        \includegraphics[width=\textwidth]{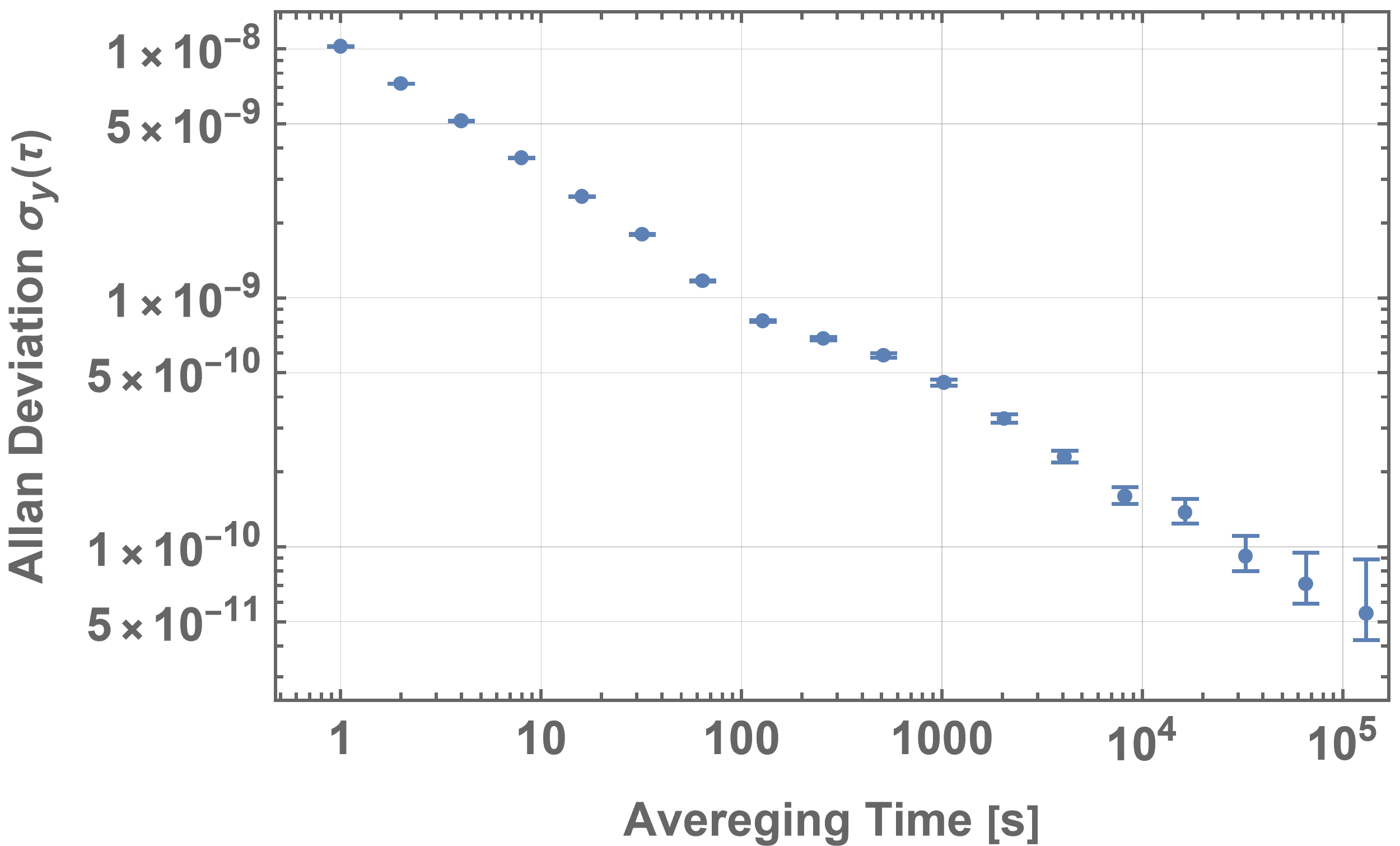}
        \caption{Frequency stability}
        \label{B_res:RW_M}
        \end{subfigure}        
    \caption{Results of the non-Markovian RW simulation described in Eq. \ref{RW_M}, with $M = 100$ without limits on phase values. a) The blue dots represent the encrypted time information signal. b) Allan deviation of the simulated time information signal depending on the averaging time with corresponding uncertainty.}
    \label{res:RW_M}
    \end{figure}

    As expected, the behavior of the induced delay shown in Fig. \ref{A_res:RW_M}  is very similar to that of a standard RW (see Fig. \ref{A_res:RW}). 
    Nevertheless, looking at the Allan deviation the correlations become evident after $M$ steps, as indicated by the change in slope in Fig. \ref{B_res:RW_M}.  
    However, any benefit of the model disappears limiting $\phi_i^{\mathrm{RW}_M}$ according to Eq. \ref{constraint} (see Fig. \ref{res:RW_Mlim}).
    \renewcommand{\thefigure}{SM5}
    \begin{figure}[h!]
    \centering
        \begin{subfigure}{0.49\textwidth}
        \includegraphics[width=\textwidth]{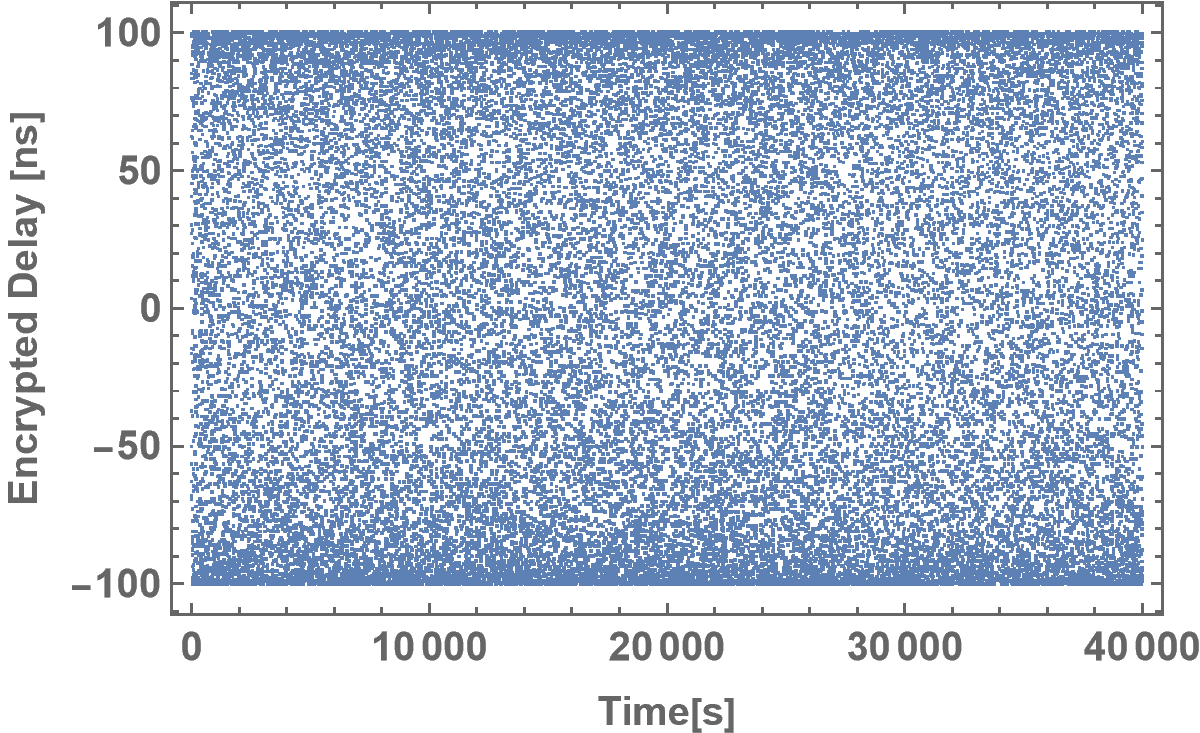}
        \caption{Time information signal}
        \label{A_res:RW_Mlim}
    \end{subfigure}
        \begin{subfigure}{0.49\textwidth}
        \includegraphics[width=\textwidth]{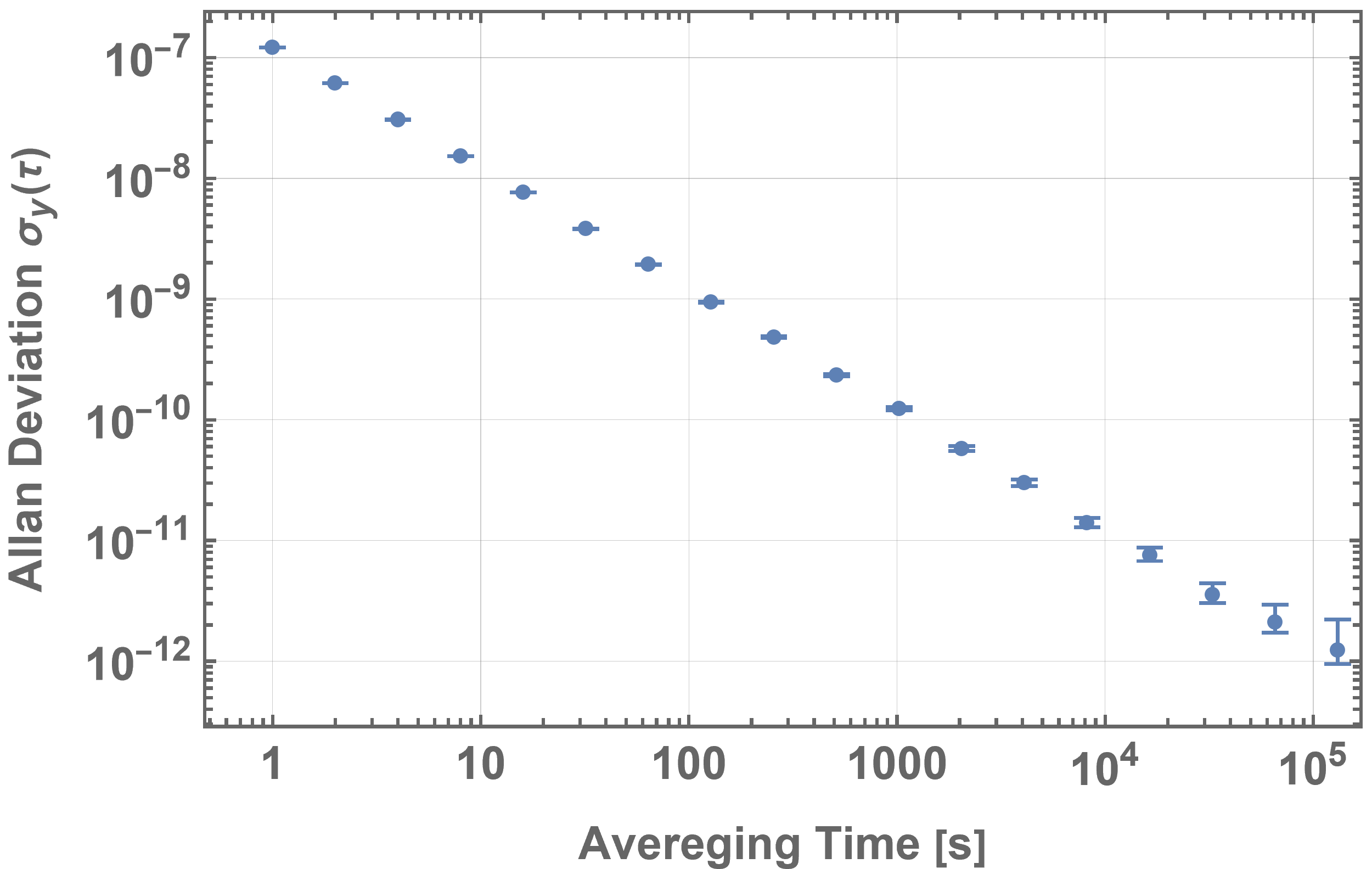}
        \caption{Frequency stability}
        \label{B_res:RW_Mlim}
        \end{subfigure}        
    \caption{Results of the non-Markovian RW simulation described in Eq. \ref{RW_M}  with $M = 100$, limiting the phase values according to Eq. \ref{constraint}. a) The blue dots represent the encrypted time information signal. b) Allan deviation of the simulated time information signal depending on the averaging time with corresponding uncertainty.}
    \label{res:RW_Mlim}
    \end{figure}
\newpage
    \subsection*{Non-Markovian Random Walk depending on last $S$ steps}
    The final phase noise model we consider is another non-Markovian extension of the random walk, starting from the key $K_3$. In this model, we increase the memory effects compared to the model described in the previous section by making the $i$-th step dependent on all of the preceding $S$ steps, with $1< S < N-1$.
    This phase evolution model is described by:
    \begin{equation}
    \phi_{i}^{\mathrm{RW}_S} = \left\{
        \begin{split}
            & \phi_{i}^{\mathrm{RW}}\quad\quad,\quad\quad i < S\\
            & \phi_{i-1}^{\mathrm{RW}_S} + \frac{\sum_{j=1}^S \Delta\phi_{i-j}^{\mathrm{RW}_S} + \frac{\left(k_{3i+1}k_{3i+2}\right)_{10}}{C}}{S} \,\,\,\,\, \mathrm{if} \,\,\,\,\,(k_{3i})\hex \geq (T_S)\hex \,\wedge\, i\geq S\\
             & \phi_{i-1}^{\mathrm{RW}_S} + \frac{\sum_{j=1}^S \Delta\phi_{i-j}^{\mathrm{RW}_S} - \frac{\left(k_{3i+1}k_{3i+2}\right)_{10}}{C} }{S} \,\,\,\,\, \mathrm{if} \,\,\,\,\,(k_{3i})\hex < (T_S)\hex \,\wedge\, i\geq S
        \end{split}
        \right.\quad\quad\quad\,\,\,
    \tag{SM5}
    \label{RW_S}
    \end{equation}
    where $i \in \left[0,\, N-1\right]$, $\phi_{i}^{\mathrm{RW}}$ is defined in Eq. \ref{RW}, and $\Delta\phi_{i-j}^{\mathrm{RW}_S} \equiv \phi_{i-j}^{\mathrm{RW}_S} - \phi_{i-j-1}^{\mathrm{RW}_S}$. Here, we select a threshold of $(T_S)\hex = 8\hex$, where we add the average of the previous $S$ increments $\Delta\phi^{\mathrm{RW}_S}$ to a balanced RW contribution (appropriately scaled by $S$).
    
    In Fig. \ref{res:RW_S}, we present the simulation of this model without any constraints on the phase $\phi_{i}^{\mathrm{RW}_S}$.
    \renewcommand{\thefigure}{SM6}
    \begin{figure}[h!]
    \centering
        \begin{subfigure}{0.49\textwidth}
        \includegraphics[width=\textwidth]{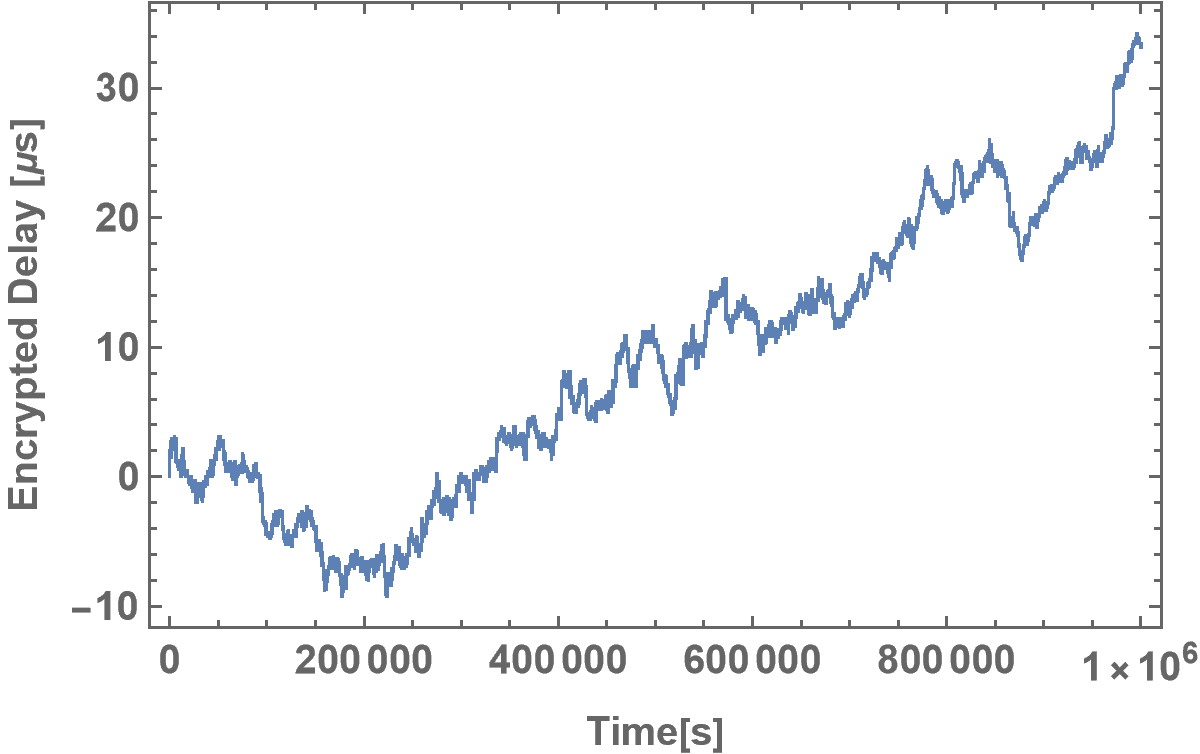}
        \caption{Time information signal}
        \label{A_res:RW_S}
    \end{subfigure}
        \begin{subfigure}{0.49\textwidth}
        \includegraphics[width=\textwidth]{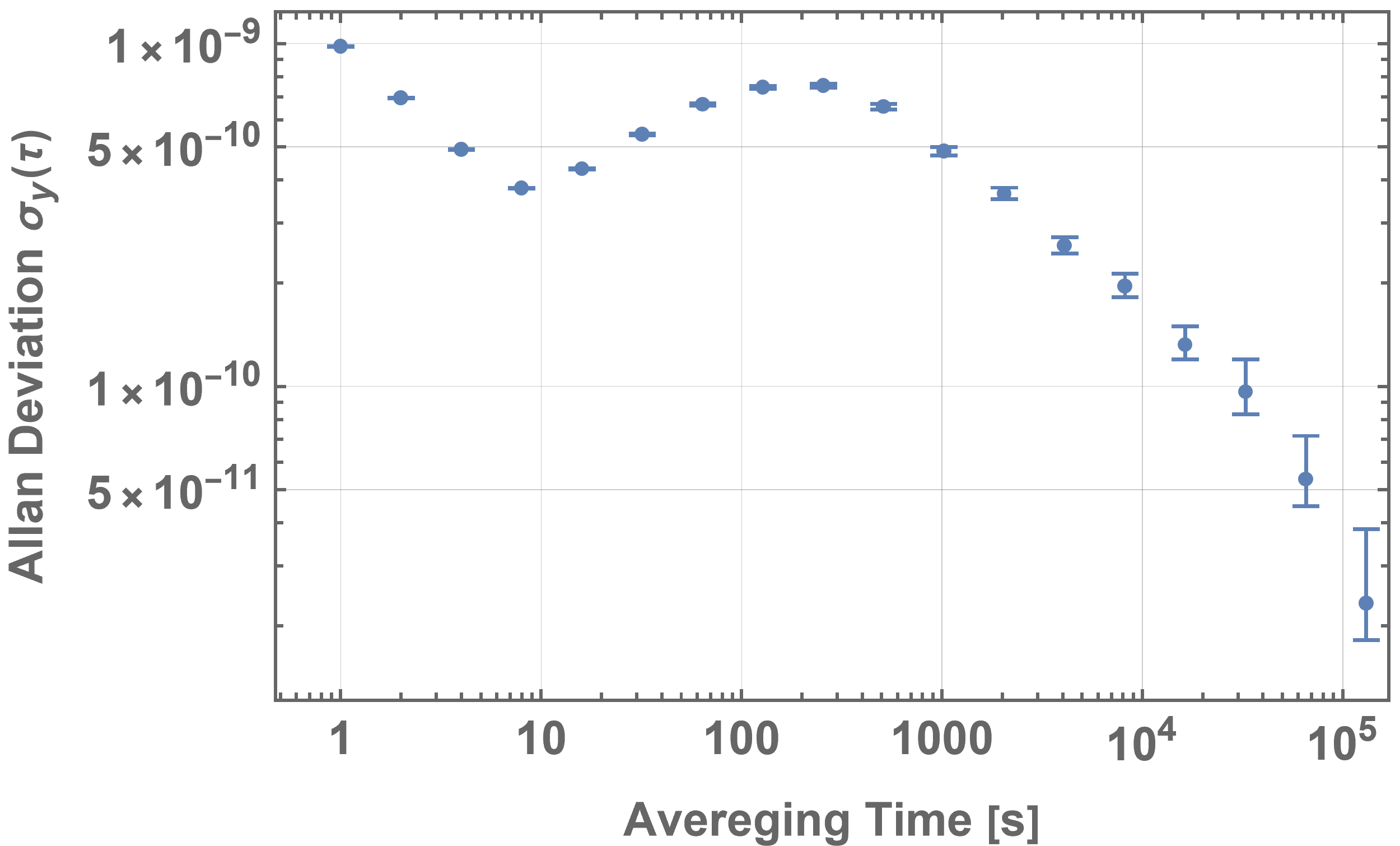}
        \caption{Frequency stability}
        \label{B_res:RW_S}
        \end{subfigure}        
    \caption{Results of the non-Markovian RW simulation described in Eq. \ref{RW_S}  with $S = 10$, without any constraint on $\phi_{i}^{\mathrm{RW}_S}$. a) The blue dots represent the encrypted time information signal. b) Allan deviation of the simulated time information signal depending on the averaging time with corresponding uncertainty.}
    \label{res:RW_S}
    \end{figure}
    In Fig. \ref{B_res:RW_S}, a clear reversal in the trend of the Allan deviation is observed after $S=10$ steps, indicating a significant advantage compared to the behavior of a standard RW. 
    However, in Fig. \ref{A_res:RW_S}, it is evident that the phase values (and thus the induced delays) diverge rapidly, making it impractical to use this noise model with the setup described in the main text. 
    For this reason, imposing a limitation as in Eq. \ref{constraint}, no advantage is gained over the white noise model presented in the main text (see Fig. \ref{res:RW_Slim}).
    \renewcommand{\thefigure}{SM7}
    \begin{figure}[h!]
    \centering
        \begin{subfigure}{0.49\textwidth}
        \includegraphics[width=\textwidth]{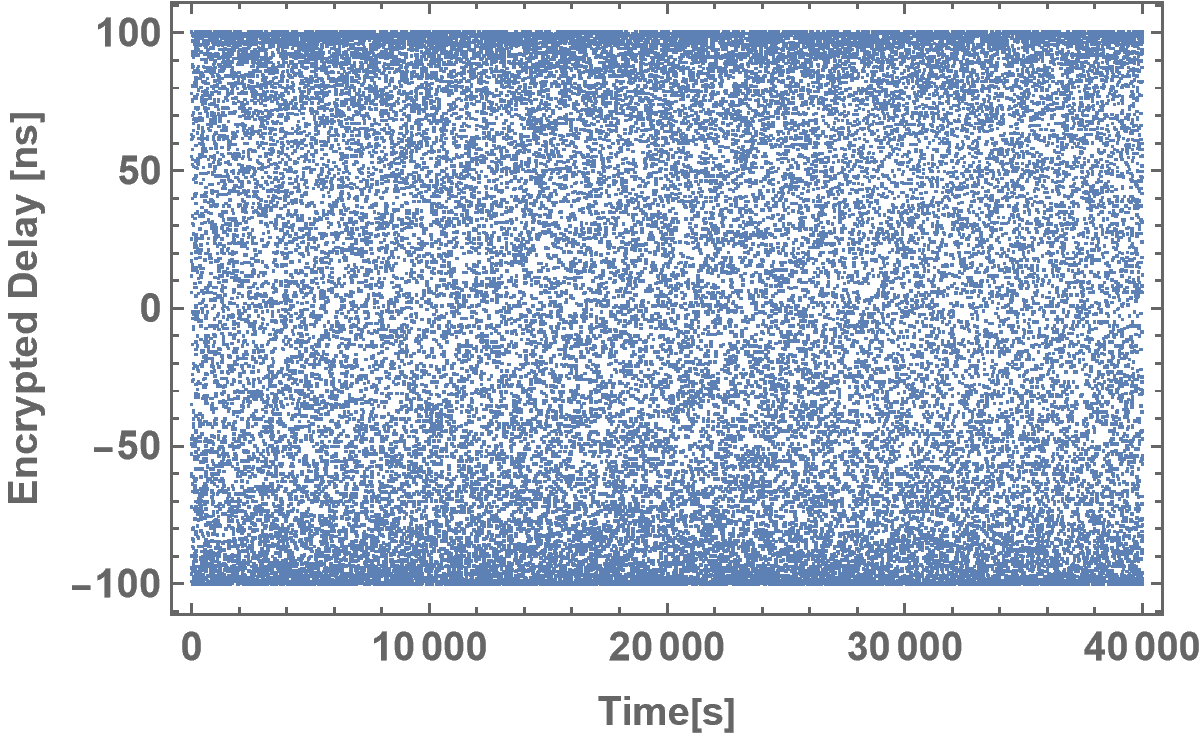}
        \caption{Time information signal}
        \label{A_res:RW_Slim}
    \end{subfigure}
        \begin{subfigure}{0.49\textwidth}
        \includegraphics[width=\textwidth]{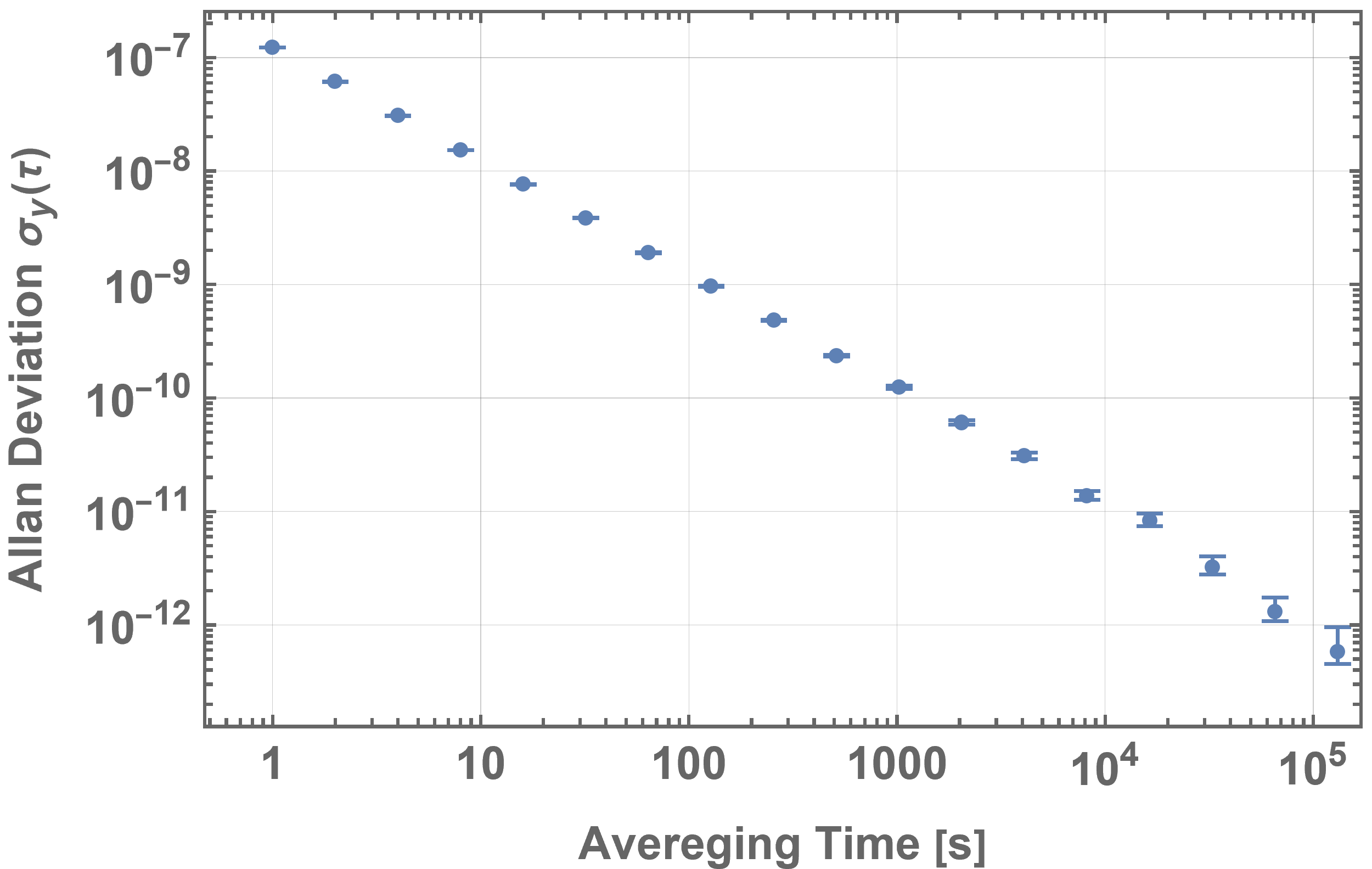}
        \caption{Frequency stability}
        \label{B_res:RW_Slim}
        \end{subfigure}        
    \caption{Results of the non-Markovian RW simulation described in Eq. \ref{RW_S}  with $S = 10$, limiting the phase values according to Eq. \ref{constraint}. a) The blue dots represent the encrypted time information signal. b) Allan deviation of the simulated time information signal depending on the averaging time with corresponding uncertainty.}
    \label{res:RW_Slim}
    \end{figure}

\section*{B) White Rabbit Protocol}

The operation of the White Rabbit Precision Time Protocol (WR-PTP) \cite{Lip11} is similar to the original PTP standard. 
Both PTP and WR-PTP rely on timestamp exchanges between the two clocks involved in synchronization, referred to as Master and Slave. 
The timestamps are typically denoted as $t_1$, $t_2$, $t_3$, and $t_4$: $t_1$ represents the time at which a synchronization packet leaves the Master, while $t_2$ indicates the time at which it is received by the Slave. 
The Slave then sends a new packet back at time $t_3$, which is received by the Master at time $t_4$. 
Since timestamps $t_1$ and $t_4$ are referenced to the Master’s clock, while $t_2$ and $t_3$ are referenced to the Slave’s clock, after a precise calibration process in which internal delays are carefully estimated, the Slave can compute the line delay $D$ and the offset $O$ used to synchronize its clock as follows:
\begin{equation}
    \begin{split}
        D &= \frac{(t_4 - t_1)- (t_3 - t_2)}{2} \\
        O &= (t_2 - t_1) - D
    \end{split}
    \tag{SM6}
    \label{WR}
\end{equation}

A key difference between PTP and WR-PTP is that, while in PTP some timestamps may be used to tune the Slave oscillator, in WR-PTP the Slave clock is always phase-locked to the Master clock via the ITU-T Synchronous Ethernet (SyncE) standard. Additionally, although some PTP server implementations can utilize SyncE to improve performance, WR-PTP achieves sub-nanosecond resolution and final uncertainty thanks to the Dual Mixer Time Difference (DMTD) detector. The DMTD precisely measures the clock transition between the received signal and the device’s internal clock. Without the dual mixer, the resolution would be limited by the internal clock’s period (e.g., for a $125\un{MHz}$ clock, the ultimate resolution would be $8\un{ns}$).
